\documentclass[journal]{IEEEtran}[10pt]
\IEEEoverridecommandlockouts
\usepackage{cite}
\usepackage{makecell}
\usepackage{amsmath,amssymb,amsfonts,amsthm}
\usepackage{algorithmic}
\usepackage{bm}
\usepackage{subfig}
\usepackage{hyperref}
\usepackage{float}  
\usepackage{graphicx}
\usepackage{makecell}
\usepackage{caption}     
\newtheorem{theorem}{Theorem}
\newtheorem{lemma}{Lemma}
\newtheorem{corollary}{Corollary}
\usepackage{textcomp}
\usepackage{epstopdf}
\usepackage{multirow}
\usepackage{xcolor}

\def\BibTeX{{\rm B\kern-.05em{\sc i\kern-.025em b}\kern-.08em
		T\kern-.1667em\lower.7ex\hbox{E}\kern-.125emX}}
\usepackage{stfloats,lipsum} 

\begin{document}
    \title{Multi-Carrier Rydberg Atomic Quantum Receivers with Enhanced Bandwidth Feature for Communication and Sensing}
	\author{Huizhi Wang,~\IEEEmembership{Graduate Student Member,~IEEE}, Tierui Gong,~\IEEEmembership{Member,~IEEE},\\ 
                Emil Björnson,~\IEEEmembership{Fellow,~IEEE}, and
			Chau Yuen,~\IEEEmembership{Fellow,~IEEE} 
    \thanks{
    This work was in part supported by MOE (Ministry of Education, Singapore), under Tier 1 Thematic Grant Award number RT12/23 023780-00001.
    		

    H. Wang is with the Graduate College, Nanyang Technological University, Singapore, under the Joint Ph.D. Programme with KTH Royal Institute of Technology, Sweden. (e-mail: HUIZHI001@e.ntu.edu.sg).
    	
    T. Gong, and C. Yuen are with School of Electrical and Electronics Engineering, Nanyang Technological University, Singapore 639798. (e-mail: tierui.gong@ntu.edu.sg, chau.yuen@ntu.edu.sg). (\emph{Corresponding author: Chau Yuen, Tierui Gong.})
    	
    E. Björnson is with Department of Computer Science, KTH Royal Institute of Technology, Stockholm, Sweden. (e-mail: emilbjo@kth.se).}
        \vspace{-1.5em}
	}
	
	\maketitle
\begin{abstract}
Rydberg atomic quantum receivers (RAQRs) have attracted significant attention in recent years due to their ultra‐high sensitivity. Although capable of precisely detecting the amplitude and phase of weak signals, conventional RAQRs face inherent limitations in accurately receiving wideband RF signals, due to the discrete nature of atomic energy levels and their intrinsic instantaneous bandwidth constraints. These limitations hinder their direct application to multi‐carrier communication and sensing. To address this issue, this paper proposes a multi‐carrier Rydberg atomic quantum receiver (MC‐RAQR) structure with five energy levels. We derive the amplitude and phase of the MC-RAQR and extract the baseband electrical signal for signal processing. In terms of multi-carrier communication and sensing, we analyze the channel capacity and accuracy of angle of arrival (AoA) and distance parameters, respectively. Numerical results validate our proposed model, showing that the MC‐RAQR can achieve up to a bandwidth of $11.7$ MHz, which is $17$-fold larger than the conventional RAQRs. As a result, the channel capacity and the resolution for multi‐target sensing are improved significantly. Specifically, the channel capacity of MC-RAQR is $110$-fold and $2.8$-fold larger than the classical RF receivers and RAQRs, respectively. For sensing performance, the \textcolor{black}{RMSE of AoA estimation for MC-RAQR exhibits $7.6$-fold reduction, compared with the conventional RAQRs. Furthermore, the RMSE of distance estimation is $634$-fold smaller than that of the root-CRB of classical RF receivers, showing the superior performance of the MC-RAQR.} This demonstrates its compatibility with waveforms such as orthogonal frequency‐division multiplexing (OFDM) and its significant advantages for multi‐carrier signal reception.
\end{abstract}
\section{introduction}
The rapid evolution of wireless networks towards the sixth generation (6G) is driven by ever increasing demands for higher data rates, massive connectivity, and ultra‐reliable low‐latency communication~\cite{6G1,6G2}, enabling diverse applications such as immersive extended reality~\cite{reality}, autonomous transportation~\cite{autonomos_vehicle}, and large‐scale Internet of Things~\cite{iot}. Achieving these ambitious goals requires not only advanced communication technologies \cite{holo1,holo2} but also breakthroughs in receiver technologies. Conventional radio frequency (RF) receivers, built on well-calibrated antennas, filters, amplifiers, and mixers~\cite{ant}, are fundamentally constrained by electronic thermal noise and the limitations of integrated circuit manufacturing processes. In this context, quantum technologies have emerged as a promising paradigm, leveraging the unique properties of quantum states to push sensing and communication capabilities beyond classical limits. Recent advances in quantum communication~\cite{gong1,gong2,cui1,sha1,qc3,chen-arxiv,xiao1}, quantum sensing~\cite{gong_doa,chief_QC_QS,chen-arxiv2}, quantum computing~\cite{qc1} and quantum network~\cite{q_network} have demonstrated unprecedented sensitivity and precision, laying the groundwork for the development of next‐generation quantum receivers capable of operating in challenging electromagnetic environments and unlocking new spectral opportunities.
In particular, highly excited Rydberg atoms have attracted extensive research interests because of their exceptionally large electric dipole moments, which enable the precise characterization of small variations in external electromagnetic fields. This remarkable property makes the Rydberg atomic quantum receiver (RAQR) a compelling candidate for achieving sensitivities on the order of ${\rm{\mu V/cm/}}\sqrt {{\rm{Hz}}}$
\cite{sensitivity}, along with an ultra‐wide reception frequency range spanning from near direct current to the terahertz (THz) regime \cite{DC_T}. Beyond its extreme sensitivity, the RAQR provides several advantages over conventional RF receivers, including SI traceable standards for electric field strength \cite{SI}, sub-wavelength imaging capabilities \cite{sub-wavelength_imaging}, and electromagnetic field detection across an ultra-wide frequency range \cite{sensing_1,sensing_2}.

As resources for wireless spectrum become increasingly scarce, there is an urgent need for RAQRs capable of receiving multi‐carrier RF signals. \textcolor{black}{However, while existing research in the physics community primarily focuses on experimental platform construction and the observation of electromagnetically induced transparency (EIT), there is a notable lack of a rigorous theoretical framework for precise signal information recovery. For practical wireless applications, it is not only necessary to detect the presence of RF fields but also essential to accurately demodulate complex communication symbols and extract sensing parameters, which remains a significant challenge for conventional resonance-based schemes.} Besides, most existing RAQRs can only operate effectively when the RF signal frequency matches a resonant transition of the Rydberg atoms, and are thus typically limited to single‐carrier, narrowband reception. Moreover, when the carrier frequency deviates from the resonance, the sensitivity of RAQRs drops sharply, rendering them ineffective for reliable RF signal detection. These shortcomings stem from the following two fundamental limitations.

Firstly, reliance on resonance-based schemes inherently restricts the range of operation frequency. Typical RAQR‐based RF field detection schemes rely on the EIT~\cite{EIT} and Autler–Townes (AT) splitting~\cite{AT}, which provides the highest sensitivity only at discrete RF transition frequencies. As a result, the RF signal can be accurately measured only within a narrow band centered on the Rydberg atomic resonance. When the RF field is detuned from this resonance frequency, the two peaks of the EIT spectrum become asymmetric, and one of them vanishes once the detuning exceeds a certain threshold~\cite{off_resonant_EIT}. This phenomenon severely limits the feasibility of receiving continuous-frequency signals with RAQRs.

Secondly, RAQRs possess an intrinsic instantaneous bandwidth determined by the atomic response time required to populate atoms into the Rydberg state. This time can be approximately expressed as $2/(\Gamma+2\gamma)$, where $\Gamma$ and $\gamma$ denote the dephasing rates of the intermediate and ground states, respectively \cite{amplitude_com}. Numerical calculations in the time domain indicate a typical value of approximately $1\mu s$ \cite{mus}, corresponding to an atomic response bandwidth on the order of $1$ MHz. When the signal bandwidth significantly exceeds this limit, the atoms do not reach a steady state within the modulation period, preventing RAQRs from fully characterizing the transmitted communication or sensing information. These inherent limitations have hindered the adoption of conventional RAQRs in scenarios demanding wideband multi-carrier reception. \textcolor{black}{On the one hand, it is fundamentally incompatible with the wideband OFDM signals in 5G/6G systems that normally needs hundreds of MHz bandwidth. On the other hand, it restricts radar range resolution to approximately $200$ meters, which means targets within $200$ m will not be separable. Such limitations render conventional 4-level RAQRs impractical for high-resolution multi-target sensing and high-speed data transmission.}

To address the aforementioned limitations, the physics community has proposed a variety of experimental schemes. For large detuning and continuous spectral bands, researchers have employed the AC Stark shift to detect off-resonant signals, at the cost of reduced sensitivity~\cite{continu1}. A two-photon AT splitting scheme has been implemented to achieve high-sensitivity detection of off-resonant signals~\cite{continu2}. Furthermore, a large-detuning Rydberg heterodyne technique has been utilized to enable digital communication with continuous carriers~\cite{continu3}. However, these approaches focus on single‐carrier signal detection and have yet to extend the bandwidth of the baseband signal. On the other hand, integrating microwave frequency comb (MFC) techniques into Rydberg reception has evolved the superheterodyne architecture from a conventional RAQR to multiple local MW fields, expanding the bandwidth to $125$ MHz~\cite{MFC}. In addition, by establishing a model for the time-dependent response of Rydberg atoms, a new receiver concept based on spatiotemporal shaping of the probe laser within a Rydberg vapor cell has been proposed to enhance the response speed of Rydberg sensors to beyond $100$ MHz~\cite{transient}. Nevertheless, this model is highly complex and does not yield a closed-form steady-state solution, which presents challenges for practical applications in communication and sensing. Recently, a novel physical mechanism has been introduced to enhance the atomic transient by generating a sideband wave in the six-wave mixing process, achieving a sensitivity of $62$ $\rm{nV}/cm/ \sqrt{Hz}$ and an instantaneous bandwidth of up to $10.2$ MHz~\cite{six_wave1,chen-arxiv3}.

To extend the capability of conventional RAQRs towards multi‐carrier signal communication and sensing over a continuous frequency range, this paper proposes a multi‐carrier Rydberg atomic quantum receiver (MC‐RAQR) structure, constructs its corresponding signal model, and performs a comprehensive performance analysis.
The main contributions are summarized as follows:
\begin{itemize}

\item \textcolor{black}{\textbf{Novel Five-Level Quantum Structure for Multi-carrier Signal Reception:} To overcome the instantaneous bandwidth bottleneck of conventional four-level RAQRs, we introduce a novel five-level quantum architecture. By incorporating an auxiliary field (AUX) and a MFC field to couple three Rydberg states, this structure fundamentally mitigates the bandwidth restriction, expanding the achievable baseband bandwidth to over $11.7$ MHz and rendering RAQRs compatible with modern communication waveforms.}

\item \textcolor{black}{\textbf{Rigorous Quantum-to-Classical Transduction Framework:} Moving beyond the traditional physics-community focus of merely detecting field presence via the EIT effect, we establish a comprehensive analytical model tailored for precise signal demodulation. We formulate the system through its corresponding Hamiltonian and solve the master equation for the density matrix under the simultaneous influence of MFC, AUX fields , and multi-carrier RF signals. Furthermore, we analyze the amplitude and phase of the output probe laser for a receiver array composed of multiple vapor cells based on the superposed incident signals. Crucially, we derive a closed-form expression for the MC-RAQR gain that explicitly links quantum-level parameters with the electrical baseband signal, thereby bridging the gap between atomic-level physical variations and receiver performance.}

\item \textcolor{black}{\textbf{Superheterodyne Multi-Carrier Signal Recovery:} To facilitate accurate phase and amplitude extraction from the superposed incident signals, we adopt a superheterodyne quantum reception scheme. Next, we derive the analytical expressions for the effective RF signal and its corresponding Rabi frequency. Furthermore, we propose two distinct MFC schemes, namely the uniform and non-uniform MFC schemes to effectively receive multi-carrier RF signals, enabling a transition from ``sensing a field" to ``recovering complex information".}

\item \textcolor{black}{\textbf{Theoretical Performance Bounds for Quantum-Enhanced Networks:} Based on the proposed five-level quantum model, we evaluate the fundamental limits for both communication and multi-target sensing tasks from a systemic perspective. For communication purpose, we formulate the achievable multi-user channel capacity as a function of the quantum MC-RAQR gain. For sensing purpose, we derive the Cramér–Rao bounds (CRBs) for angle-of-arrival (AoA) and range estimation using the multiple signal classification (MUSIC) algorithm~\cite{music}. Extensive simulations validate the feasibility of the architecture, demonstrating super-resolution sensing and enhanced capacity, and enabling a direct theoretical comparison between the MC-RAQR, the conventional RAQRs and classical RF receivers.}
\end{itemize}

\textit{Organization and Notations}: 
The rest of the article is organized as follows: In Section \ref{system}, we propose a five-level MC-RAQR and construct the signal model. In Section \ref{orsp}, we present the signal processing procedure and the output signal. In Section \ref{performance_analysis}, we provide comprehensive analysis of wireless communication and sensing tasks. Then we present our simulation results in Section \ref{simulation}, and finally conclude in Section \ref{conclusion}. 
We use the following notations: $\dot{ \bm{\rho}}$ is the differential of $\bm{\rho}$ with respect to time; $[{\bm{\rho}},{\bm{H}}] = \bm{\rho}\bm{H}-\bm{H}\bm{\rho}$ represents the commutator; $\mathcal{R} \{ \cdot \}$ and $\mathcal{I} \{ \cdot \}$ take the real and imaginary parts of a complex number; ${\rm{diag}}\{ \cdot \}$ is the diagonal matrix; $\hbar$ is the reduced Planck constant; $c$ and $\epsilon_0$ are the speed of light in free space and the vacuum permittivity; $\eta$ and $q$ are the quantum efficiency and the elementary charge, respectively. 

\section{System model of MC-RAQR}\label{system}
	\begin{figure}[ht]
	\subfloat[\textcolor{black}{The structure of MC-RAQRs.}]{
			\includegraphics[width=0.48\textwidth]{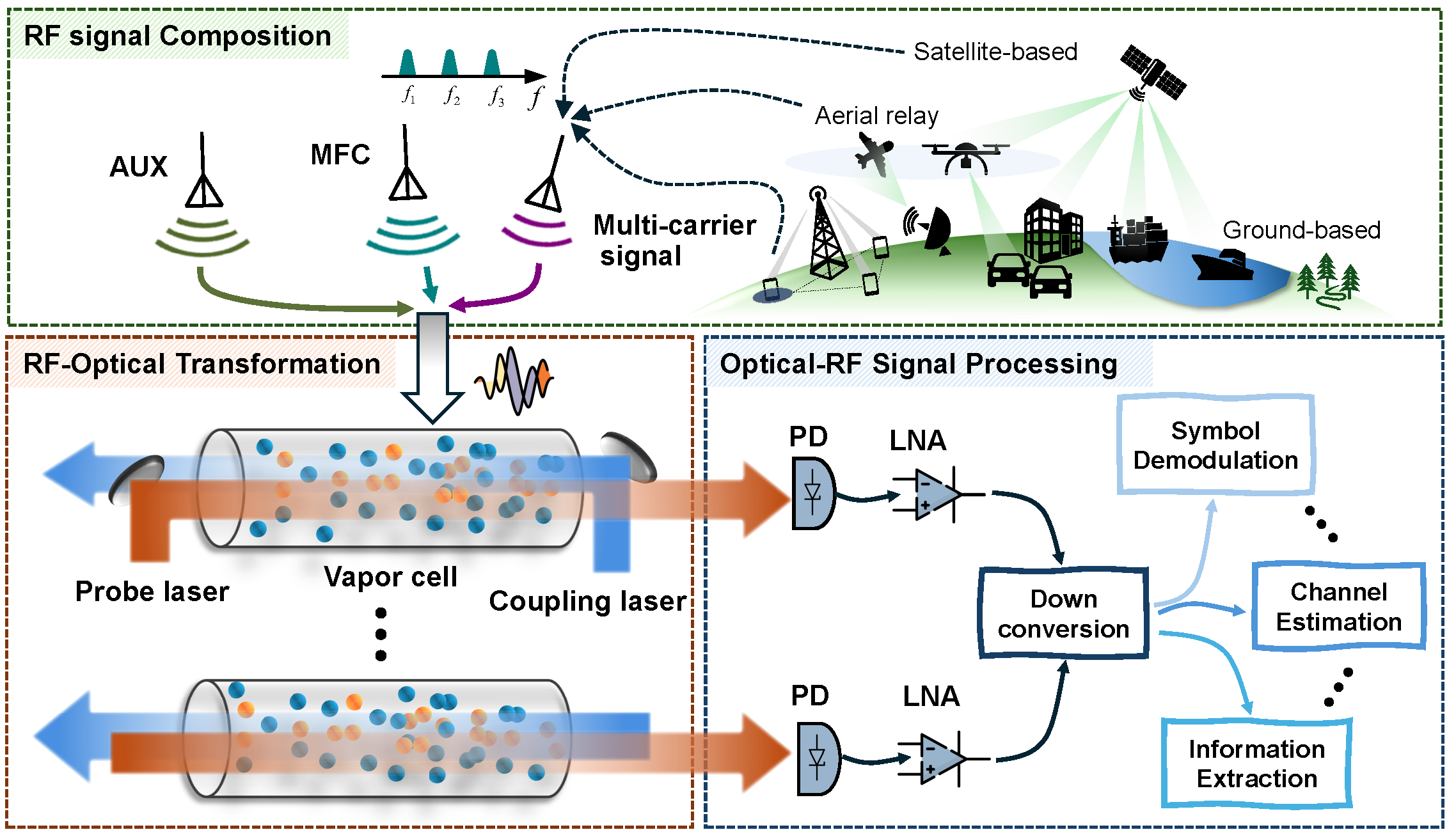}
			\label{flow}
		}\\
		\subfloat[Five-level diagram.]{
			\includegraphics[width=0.48\textwidth]{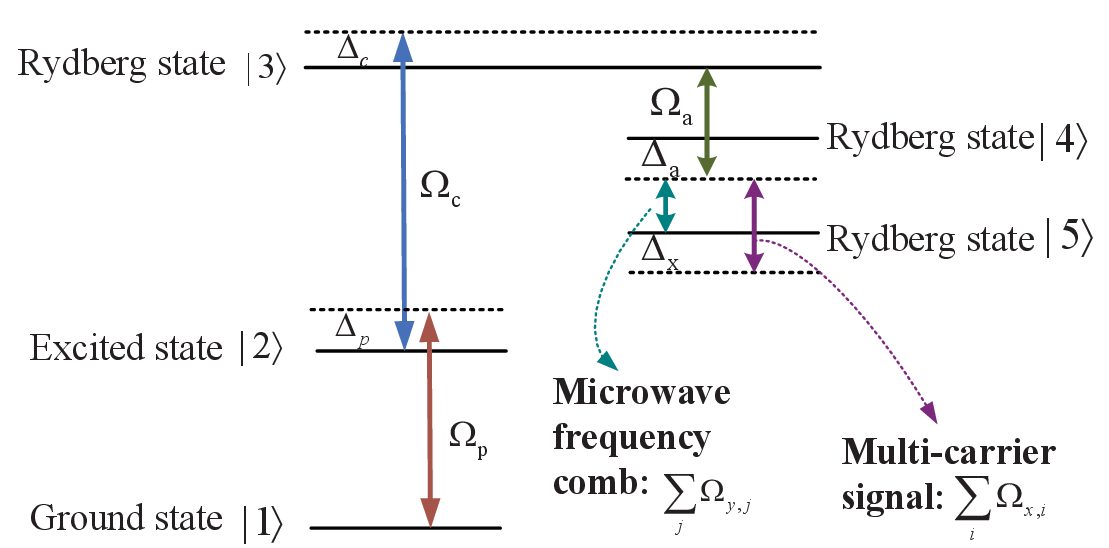}
			\label{five-level}
		}
		\caption{\textcolor{black}{The system architecture and physical model of the proposed MC-RAQR scheme. (a) The system structure of MC-RAQRs, including the RF signal composition, RF-optical transformation, and the optical-RF signal processing. (b) The five-level atomic energy diagram of Cs atoms interacting with the probe laser ($\Omega_p$), coupling laser ($\Omega_c$), AUX, MFC, and the incident multi-carrier RF signal.}}
		\label{pattern}
	\vspace{-2ex}
	\end{figure}
\textcolor{black}{Fig. \ref{pattern} (a) illustrates the structure of the proposed MC-RAQR, which primarily consists of three components: the RF signal composition, RF-optical transformation composed of $M$ vapor cells, and the optical-RF signal processing. In the following, we establish the system models for these three components, respectively.}
\subsection{RF Signal Composition}\label{RF}
\textcolor{black}{In conventional RAQRs, the incident RF signal is typically modeled as a superposition of the desired RF signal and a local oscillator (LO), where the employed LO is conventionally a single-frequency field. To enable the efficient reception of multi-carrier wireless signals, this paper employs a multi-frequency MFC signal and introduces an AUX field. This novel design enables continuous signal reception with enhanced bandwidth across the frequency domain.}
The multi-carrier signal received at the $m$-th Rydberg sensor can be expressed as
\begin{equation}
	\setlength\abovedisplayskip{2.7pt}
	\setlength\belowdisplayskip{2.7pt}
	{x_m}(t) = \sum\limits_{i = 0}^{N-1} {\sqrt {2{\mathcal{P}_{x,i,m}}} \cos (2\pi {f_{x,i}}t + {\theta _{i,m}})}, 
\end{equation} 
where $\mathcal{P}_{x,i,m}$ and $\theta_{i,m}$ denote the power and phase of the $i$-th frequency band at the $m$-th sensor, respectively. $f_{x,i}$ denotes the frequency of the $i$-th subcarrier. Note that $x_m(t)$ can be also expressed using the complex baseband signal as
\begin{equation}
	\setlength\abovedisplayskip{2.7pt}
	\setlength\belowdisplayskip{2.7pt}
	{x_m}(t) = \sqrt 2 \mathcal{R}\left\{ {\sum\limits_{i = 0}^{N-1} {{x_{i,m}}(t){e^{j2\pi {f_{x,i}}t}}} } \right\},
	\label{xm}
\end{equation}
where ${x_{i,m}}(t) = \sqrt {{\mathcal{P}_{x,i,m}}} {e^{j{\theta _{i,m}}}}$ is the equivalent baseband signal of the $i$-th carrier.

Similarly, the MFC signal at the $m$-th sensor can be formulated as follows 
\begin{equation}
	\setlength\abovedisplayskip{2.7pt}
	\setlength\belowdisplayskip{2.7pt}
	\begin{aligned}
		{y_m}(t) &= \sum\limits_{j = 1}^B {\sqrt {2{\mathcal{P}_{y,j,m}}} \cos (2\pi {f_{y,j,m}}t + {\varphi _{j,m}})} \\
		&= \sqrt 2 \mathcal R \Bigg\{ {\sum\limits_{j = 1}^B {{y_{j,m}}(t){e^{j2\pi {f_{y,j,m}}t}}} } \Bigg\},
	\end{aligned}
	\label{ym}
\end{equation}
where $B$ is the number of MFC comb lines. $\mathcal{P}_{y,j,m}, \varphi_{j,m}$ and $f_{y,j,m}$ denote the power, phase and frequency of the $j$-th comb line at the $m$-th sensor, respectively; ${y_{j,m}}(t) = \sqrt {{\mathcal{P}_{y,j,m}}} {e^{j{\varphi _{j,m}}}}$ represents the baseband signal of MFC. Without loss of generality, we assume the power of all comb lines are identical, i.e., ${\mathcal{P}_{y,j,m}} = \frac{ {\mathcal{P}_{y,m}}}{B},\forall j.$

The RF signal to be detected is the superposition of the multi-carrier signal and MFC, which can be written as
\begin{equation}
	\setlength\abovedisplayskip{2.7pt}
	\setlength\belowdisplayskip{2.7pt}
	\hspace{-2.2ex}
	\begin{aligned}
{z_m}(t) = {x_m}(t) + {y_m}(t) = \sum\limits_l {\sqrt {2{{\cal P}_{z,l,m}}} \cos (2\pi {f_{z,l,m}}t + {\phi _{z,l,m}})} ,
	\end{aligned}
	\label{zmt}
\end{equation}
where $\mathcal{P}_{z,l,m}, \phi_{z,l,m}$ and $f_{z,l,m}$ denote the power, phase and frequency of the $l$-th component of RF signal at the $m$-th sensor, respectively. \textcolor{black}{Note that while the superimposed signal in (\ref{zmt}) is expressed in terms of the power $\mathcal{P}_{z,l,m}$, the subsequent RF-optical transformation fundamentally relies on the RF electric field amplitude to modulate the probe laser. To facilitate the analysis of this physical process, we define the corresponding signal amplitude as $U_{z,m} = \sqrt{\frac{2\mathcal{P}_{z,m}}{A_e c \epsilon_0}}$, where $A_e$ denotes the effective RAQR aperture. The expression for the combined amplitude $U_{z,m}$ is explicitly derived below.}

\begin{lemma}\label{Uzm}
	The amplitude of the RF signal to be detected can be approximately formulated  as
	\begin{equation}
		\setlength\abovedisplayskip{2.7pt}
		\setlength\belowdisplayskip{2.7pt}
		{U_{z,m}} \approx {U_{y,m}} + \frac{1}{{\sqrt B }}\sum\limits_{i = 0}^{N-1} {{U_{x,i,m}}\cos (2\pi \Delta {f_{ni,m}}t + {\Delta \phi_{ni,m}})}, 
		\label{uzm}
	\end{equation}
	where ${U_{y,m}} = \sqrt {\frac{{2{\mathcal{P}_{y,m}}}}{{{A_e}c{\epsilon_0}}}}$ and ${U_{x,i,m}} = \sqrt {\frac{{2{\mathcal{P}_{x,i,m}}}}{{{A_e}c{\epsilon_0}}}}$ denote the amplitude of the MFC and multi-carrier signal, respectively. $\Delta {f_{ni,m}} = {f_{x,i}} - {f_{y,ni,m}}$ and $\Delta {\phi _{ni,m}} = {\theta _{i,m}} - {\varphi _{ni,m}}$ represent the frequency and phase of the intermediate frequency (IF) signal between the $i$-th frequency band the its nearest MFC comb line at the $m$-th sensor, respectively.
	\begin{IEEEproof}
		Please refer to Appendix A.
	\end{IEEEproof}
\end{lemma}
Lemma \ref{Uzm} shows that the amplitude of signal impinging on the $m$-th sensor is closely related to the amplitude of MFC field and multi-carrier signal. Besides, the frequency and phase of the beat signal between each carrier and its nearest MFC comb line is also included in $U_{z,m}$, which can be extracted and recovered by proper signal processing.

\subsection{RF-Optical Transformation}
\textcolor{black}{To receive the superimposed RF signals (\ref{uzm}), we employ $M$ vapor cells of length $L$. Each vapor cell is filled with alkali metal atoms (e.g., cesium, Cs) and serves as an equivalent quantum sensor. Additionally, all $M$ vapor cells form a uniform linear array (ULA) with a sensor element spacing of $d$. When the probe and coupling lasers are spatially overlapped and counter-propagating through each vapor cell, the atoms within this region are excited to Rydberg states, where their atomic responses are carried to the probe laser. When the above-mentioned RF signal superimposition is imposed, the atomic response and the probe laser are influenced accordingly. The influence of this interaction is read out by photodetectors (PDs), which convert the optical output of the probe laser into an electrical signal for subsequent processing. }
	
\textcolor{black}{To describe the reactions occurring within each atom more clearly, we utilize an energy level diagram as illustrated in Fig. \ref{pattern} (b). Specifically, each target atom absorb energy from the applied lasers or external electromagnetic wave signals and then transit to higher energy levels. Based on this principle, each target atom first transits from the ground state $|1\rangle$ to the excited state $|2\rangle$ under the illumination of the probe laser, then transits from $|2\rangle$ to the Rydberg state $|3\rangle$ under the illumination of the coupling laser. At this highly excited state, the atoms become exceptionally sensitive to variations in external electric fields. Driven by the newly introduced AUX field, each target atom transits to Rydberg state $|4\rangle$, and ultimately arrives at the Rydberg state $|5\rangle$ under the joint driving force of the MFC field and multi-carrier signal. To mathematically characterize the impact of these optical and electrical signals on each target atom, we utilize the Rabi frequency $\Omega_o$ and the detuning $\Delta_o$ to embed the amplitude and frequency information, respectively, where the subscript $o \in \{p, c, a, z\}$ corresponds to the probe laser, coupling laser, AUX field, and the superimposed RF signal, respectively. Physically, the transition of the proposed five-level system can be described using a Hamiltonian \cite{three_aux_model}, in which the diagonal elements represent the detuning of each state, while the off-diagonal elements indicate the coupling between different energy levels. The Hamiltonian matrix of the $m$-th vapor cell in MC-RAQR can be modeled as
\begin{equation}
	\setlength\abovedisplayskip{2.7pt}
	\setlength\belowdisplayskip{2.7pt}
	{\bf{H}}_m(t) = \frac{\hbar }{2}\left[ {\begin{array}{*{20}{c}}
			0&{{\Omega _p}}&0&0&0\\
			{{\Omega _p}}&A&{{\Omega _c}}&0&0\\
			0&{{\Omega _c}}&B&{{\Omega _{{\rm{a}}}}}&0\\
			0&0&{{\Omega _{{\rm{a}}}}}&C&{{\Omega _{z,m}}}\\
			0&0&0&{{\Omega _{z,m}}}&D
	\end{array}} \right],
\end{equation}
where $A =  - 2{\Delta _p},B =  - 2({\Delta _p} + {\Delta _c}),C =  - 2({\Delta _p} + {\Delta _c} - {\Delta _a}),D =  - 2({\Delta _p} + {\Delta _c} - {\Delta _a} - {\Delta _{x}}).$ ${\Omega _{z,m}}$ is the Rabi frequency of the superimposed signal $z_m(t)$, which can be obtained by substituting $\Omega_{t,m}=\frac{\mu_{45}}{\hbar}U_{t,m}, t=\{x,y,z\}$ into (\ref{uzm}), where $\mu_{45}$ denotes the dipole moment between the Rydberg states $|4\rangle$ and $|5\rangle$. After some simple derivations, we obtain the expression of ${\Omega _{z,m}}$ as follows 
\begin{equation}
	\setlength\abovedisplayskip{2.7pt}
	\setlength\belowdisplayskip{2.7pt}
	{\Omega _{z,m}} \approx {\Omega _{y,m}} + \frac{1}{{\sqrt B }}\sum\limits_{i = 0}^{N-1} {{\Omega _{x,i,m}}\cos (2\pi \Delta {f_{ni,m}}t + \Delta {\phi _{ni,m}})}.
\end{equation}
Based on the Hamiltonian ${\bf{H}}_m(t)$, we employ the Lindblad master equation to model the dynamic evolution process of the quantum system \cite{master_equation}
\begin{equation}
	\setlength\abovedisplayskip{2.7pt}
	\setlength\belowdisplayskip{2.7pt}
	\label{master}
	\dot{ \bm{\rho}}_m  = \frac{i}{\hbar }[{\bm{\rho}}_m ,{\bf{H}}_m(t)] + \bm D(\bm{\rho}_m),
\end{equation}
where ${ \bm{\rho}}_m$ is the density matrix. Specifically, the diagonal element ${\bm{\rho}}_{ii,m}$ denotes the population probability, representing the likelihood of an atom occupying the specific energy state $|i\rangle$. Conversely, the off-diagonal element ${\bm{\rho}}_{ij,m}$ denotes the quantum coherence between state $|i\rangle$ and state $|j\rangle$, which characterizes the phase synchronization and transition dynamics driven by the external optical and RF fields. $\bm D(\bm{\rho}_m)$ represents the Lindblad operator accounting for relaxation and decay processes. Since the spontaneous emission associated with $|3\rangle$, $|4\rangle$ and $|5\rangle$ are comparatively small, they can be reasonably ignored and $\bm D({\bm\rho}_m )$ can be simplified as 
\begin{equation}
	\setlength\abovedisplayskip{2.7pt}
	\setlength\belowdisplayskip{2.7pt}
	\small
	\bm D({\bm\rho} ) = \left[ {\begin{array}{*{20}{c}}
			{{\gamma _2}{\rho _{22}}}&{ - \frac{{{\gamma _2}}}{2}{\rho _{12}}}&0&0&0\\
			{ - \frac{{{\gamma _2}}}{2}{\rho _{21}}}&{ - {\gamma _2}{\rho _{22}}}&{ - \frac{{{\gamma _2}}}{2}{\rho _{23}}}&{ - \frac{{{\gamma _2}}}{2}{\rho _{24}}}&{ - \frac{{{\gamma _2}}}{2}{\rho _{25}}}\\
			0&{ - \frac{{{\gamma _2}}}{2}{\rho _{32}}}&0&0&0\\
			0&{ - \frac{{{\gamma _2}}}{2}{\rho _{42}}}&0&0&0\\
			0&{ - \frac{{{\gamma _2}}}{2}{\rho _{52}}}&0&0&0
	\end{array}} \right],
\end{equation}
where $\gamma_2$ is the spontaneous decay rate of $|2\rangle$.
}

To obtain useful insights, we assume the probe laser is perfectly resonant, i.e., $\Delta_p=0$. By further solving (\ref{master}) based on the steady-state condition, the density matrix can be obtained. \textcolor{black}{More particularly, we focus on the element $\rho_{21,m}$ as it directly governs the microscopic quantum coherence between the ground state $|1\rangle$ and the excited state $|2\rangle$, which ultimately determines the optical susceptibility and the transmission properties of the probe laser. In the steady-state condition, we analytically obtain the closed-form expression for $\rho_{21,m}$ as}
\begin{equation}
		\setlength\abovedisplayskip{2.7pt}
	\setlength\belowdisplayskip{2.7pt}
\label{rho21}
{\rho _{21,m}} = \frac{{{{\bar o}_1}\Omega _{z,m}^4 + {{\bar o}_2}\Omega _{z,m}^2 + {{\bar o}_3} + j({{\bar o}_4}\Omega _{z,m}^4 + {{\bar o}_5}\Omega _{z,m}^2 + {{\bar o}_6})}}{{{\underline{o}_1}\Omega _{z,m}^4 + {\underline{o}_2}\Omega _{z,m}^2 + {\underline{o}_3}}},
\end{equation}
where ${\bar o}_{1},{\bar o}_2, {\bar o}_3, {\bar o}_4, {\bar o}_5, {\bar o}_6, \underline{o}_1, \underline{o}_2, \underline{o}_3$ are coefficients given in Appendix B. Let $\chi_m(\Omega)=C\rho_{21,m}(\Omega)$ be the atomic susceptibility and $C = \frac{{ - 2{N_0}\mu _{12}^2}}{{{\epsilon_0}\hbar {\Omega _p}}}$, where $N_0$ is the atomic density in the vapor cell. $\mu_{12}$ represents the dipole moment of the transition between states $|1\rangle$ and $|2\rangle$. The amplitude and phase of the probe laser can be obtained as
\begin{equation}
		\setlength\abovedisplayskip{2.7pt}
	\setlength\belowdisplayskip{2.7pt}
	\label{ampli_phase}
\begin{aligned}
	{U_{p,m}}({\Omega_{z,m}}) &= {U_{o,m}}{e^{ - {k_p}L\mathcal{I}[{\chi _m}({\Omega_{z,m}})]}}, \\
	{\phi _{p,m}}({\Omega_{z,m}}) &= {\phi _{0,m}} + {k_p}L\mathcal{R}[\chi_m ({\Omega_{z,m}})],
\end{aligned}
\end{equation}
where $U_{0,m}$ and $\phi_{0,m}$ denote the amplitude and phase of the input probe laser, respectively; ${k_p} = \frac{{2\pi }}{{{\lambda _p}}}$ is the wavenumber of probe laser with $\lambda_p$ being the wavelength. Therefore, the probe laser at the output of the $m$-th sensor can be written as
\begin{equation}
		\setlength\abovedisplayskip{2.7pt}
	\setlength\belowdisplayskip{2.7pt}
\begin{aligned}
	{P_m}({\Omega_{z,m}},t) &= \sqrt {2{\mathcal{P}_m}({\Omega_{z,m}})} \cos (2\pi {f_p}t + {\phi _{p,m}}({\Omega_{z,m}}))\\
	&= \sqrt 2 \mathcal{R}\{ {P_{b,m}}({\Omega_{z,m}},t){e^{2\pi {f_p}t}}\}, 
\end{aligned}
\end{equation}
where ${P_{b,m}}(\Omega_{z,m},t) = \sqrt {{\mathcal{P}_m}({\Omega_{z,m}})} {e^{j{\phi _{p,m}}({\Omega_{z,m}})}} = \sqrt {\frac{{\pi c{\epsilon_0}}}{{8\ln2}}} {F_p}\left| {{U_{p,m}}({\Omega_{z,m}})} \right|{e^{j{\phi _{p,m}}({\Omega_{z,m}})}}$ denotes the equivalent baseband signal of the output probe laser and $F_p$ represents the full width at half maximum (FWHM) of the probe laser.

\section{Optical-RF signal processing}\label{orsp}
\begin{figure}[t!]
	\centering
	\includegraphics[width=0.49\textwidth]{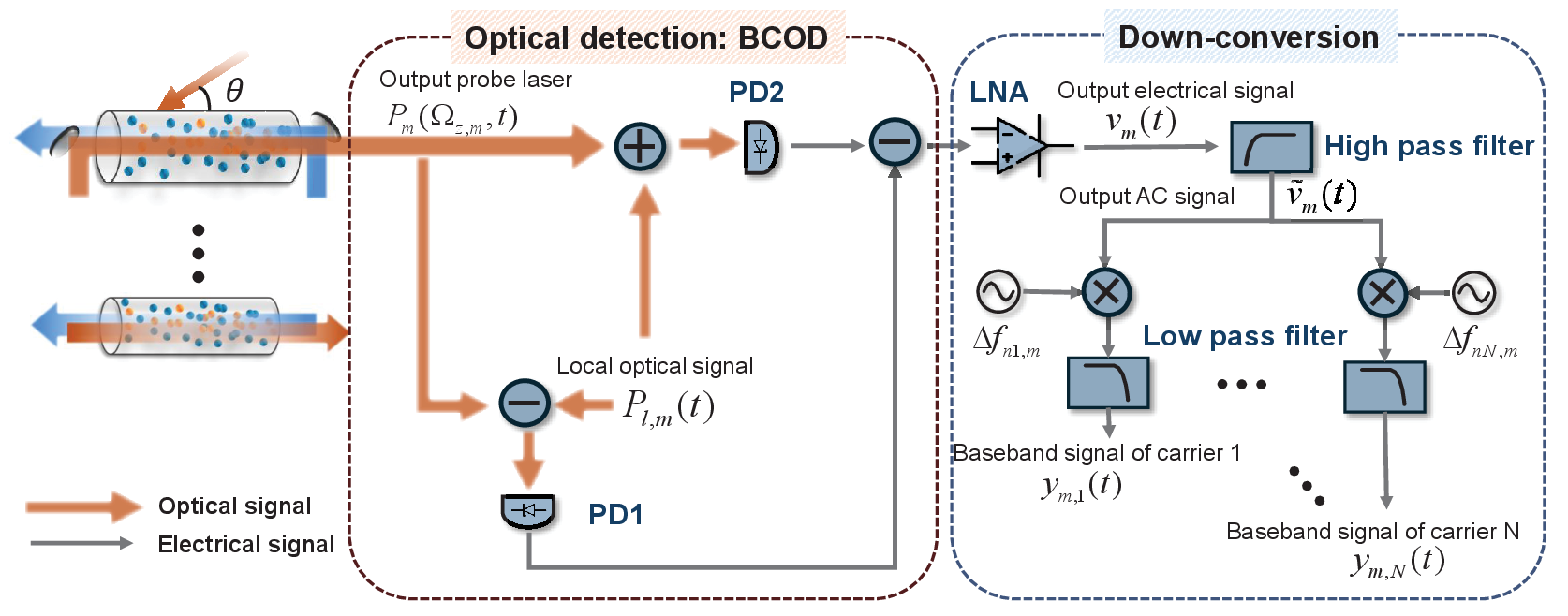}
	\caption{\textcolor{black}{Block diagram of the baseband optical-RF signal processing module, where the BCOD scheme is used to extract the alternating current (AC) signal, followed by down-conversion and low-pass filtering to separate and extract the baseband signals for individual subcarriers.}}
	\label{oddc}
    \vspace{-1.6em}
\end{figure}


To extract the impinging multi-carrier RF signal from the optical probe laser, we need to transform the signal into electrical signal through photodetection. In this paper, we employ the balanced coherent optical detection (BCOD) scheme, as shown in Fig. \ref{oddc}. In particular, one local optical signal ${P_{l,m}}(t) = \sqrt {2{\mathcal{P}_{l,m}}} \cos (2\pi {f_p}t + {\phi _{l,m}})$ is used to suppress the thermal noise generated by the electronic components, and the corresponding baseband signal is expressed as ${P_{lb,m}} = \sqrt {{\mathcal{P}_{l,m}}} {e^{j{\phi _{l,m}}}}$. The output of BCOD is given by
\begin{equation}
		\setlength\abovedisplayskip{2.7pt}
	\setlength\belowdisplayskip{2.7pt}
	\label{output_PD}
\begin{aligned}
	&{v_m}(\Omega_{z,m},t) = \sqrt {G_{\text{LNA}}} \alpha [{P_{lb,m}}(t)P_{b,m}^*({\Omega_{z,m}},t)\\ 
	&\qquad \qquad \qquad \quad + P_{lb,m}^*(t){P_{b,m}}({\Omega_{z,m}},t)]\\
	&= 2\sqrt {G_{\text{LNA}}} \alpha \sqrt {{\mathcal{P}_{l,m}}{\mathcal{P}_m}({\Omega_{z,m}})} \cos ({\phi _{l,m}} - {\phi _{p,m}}({\Omega_{z,m}})),
\end{aligned}
\end{equation}
where ${G_{\text{LNA}}}$ is the gain of the low noise amplifier (LNA) and $\alpha  = \frac{{\eta q}}{{\hbar {\omega _p}}}$ is the PD's responsivity. In the following, we retain the first-order term of $v_m(\Omega_{z,m},t)$ at $\Omega_{z,m}=\Omega_{y,m}$ by ignoring the high-order terms for gaining deeper insights. 
\begin{lemma}\label{opd}
The output of PD for the $m$-th sensor can be expressed in the form of a direct current (DC) plus an AC as 
\begin{align}
		\setlength\abovedisplayskip{2.7pt}
	\setlength\belowdisplayskip{2.7pt}
	\small
\begin{aligned}
	{v_m}(t) &\approx 2\sqrt {G_{\text{LNA}}} \alpha \sqrt {{\mathcal{P}_{l,m}}{\mathcal{P}_m}({\Omega_{y,m}})} [\cos ({\phi _{l,m}} - {\phi _{p,m}}({\Omega_{y,m}}))\\
	&+ \sqrt{\frac{2}{A_ec\epsilon_0}}{A_m}({\phi _{l,m}},{\Omega _p},{\Omega _c},{\Omega _a},{\Omega _{y,m}},{\Delta _a},B)\\
	&\times\sum\limits_{i = 0}^{N-1} {\sqrt{\mathcal{P}_{x,i,m}}}  \cos (2\pi \Delta {f_{ni,m}}t + \Delta {\phi _{ni,m}})],
\end{aligned}
\end{align}
where we have 
\begin{equation}
	\setlength\abovedisplayskip{2.7pt}
\setlength\belowdisplayskip{2.7pt}
\begin{aligned}
	\small
	\begin{aligned}
		\label{parameter_theo2}
		&{A_m} = \frac{{{\tilde A}_m}\pi d{\mu _{45}}}{{{\lambda _p}\sqrt B \hbar }}\cos ({\phi _{l,m}} - {\phi _{p,m}}({\Omega_{y,m}}) + {\psi _{p,m}}({\Omega_{y,m}})),\\
		&{{\tilde A}_m}= \sqrt {{\mathcal{I}^2}\{ {\chi _m}^\prime ({\Omega_{y,m}})\}  + {\mathcal{R}^2}\{ {\chi _m}^\prime ({\Omega_{y,m}})\} },\\
		&{\psi _{p,m}} = \arccos \frac{{\mathcal{I}\{ {\chi _m}^\prime ({\Omega_{y,m}})\} }}{{{{\tilde A}_m}}},\\
		&\mathcal{R}\{ {\chi _m}^\prime ({\Omega_{y,m}})\}  = C\frac{{4{\overline{o}_1}{\Omega_{e,m}^3} + 2{\overline{o}_2}{\Omega_{e,m}}}}{{{\underline{o}_1}\Omega_{e,m}^4 + {\underline{o}_2}\Omega_{e,m}^2 + {\underline{o}_3}}}\nonumber\\
		&~~~~~~~- C\frac{{({\overline{o}_1}{\Omega_{e,m}^4} + {\overline{o}_2}{\Omega_{e,m}^2} + {\overline{o}_3})(4{\underline{o}_1}{\Omega_{e,m}^3} + 2{\underline{o}_2}\Omega_{e,m})}}{{{{({\underline{o}_1}\Omega_{e,m}^4 + {\underline{o}_2}\Omega_{e,m}^2 + {\underline{o}_3})}^2}}},\\
		&\mathcal{I}\{ {\chi _m}^\prime ({\Omega_{y,m}})\}  = -C\frac{{4{\overline{o}_4}{\Omega_{e,m}^3} + 2{\overline{o}_5}{\Omega_{e,m}}}}{{{\underline{o}_1}\Omega_{e,m}^4 + {\underline{o}_2}\Omega_{e,m}^2 + {\underline{o}_3}}} \nonumber\\
		&~~~~~~~+C\frac{{({\overline{o}_4}{\Omega_{e,m}^4} + {\overline{o}_5}{\Omega_{e,m}^2} + {\overline{o}_6})(4{\underline{o}_1}{\Omega_{e,m}^3} + 2{\underline{o}_2}\Omega_{e,m})}}{{{{({\underline{o}_1}\Omega_{e,m}^4 + {\underline{o}_2}\Omega_{e,m}^2 + {\underline{o}_3})}^2}}}.
	\end{aligned}
\end{aligned}
\end{equation}
\begin{IEEEproof}
Please refer to Appendix C.
\end{IEEEproof}
\end{lemma}
Lemma \ref{opd} shows that $v_m(t)$ contains the information about the amplitude of impinging RF signal, and the frequency and phase of the beat signal. In the following, we remove the RF-independent DC component of $v_m(t)$ by passing through a \textcolor{black}{high-pass filter (HPF)} and only output the RF-dependent AC component as
\begin{equation}
	\setlength\abovedisplayskip{2.7pt}
	\setlength\belowdisplayskip{2.7pt}
\label{vmtilde}
\begin{aligned}
	{{\tilde v}_m}(t) = \rho_m\sum\limits_{i = 0}^{N-1} {\sqrt{\mathcal{P}_{x,i,m}}} \cos (2\pi \Delta {f_{ni,m}}t + \Delta {\phi _{ni,m}}),
\end{aligned}
\end{equation}  
where we have ${\rho _m} = 2\alpha \sqrt {\frac{{2G{\mathcal{P}_{l,m}}{\mathcal{P}_m}({\Omega_{y,m}})}}{{{A_e}c{\epsilon_0}}}} {A_m}.$

\subsection{MFC Schemes and Baseband Signal}

Furthermore, we assume different band signal are temporally orthogonal, so the output signal $\tilde{v}_m$ for each carrier $i$ can be independently recovered by demodulating with different carrier frequencies, as shown in Fig. \ref{oddc}. In practice, the orthogonality of IF frequencies may not always hold and several subcarriers may lead to the same IF, rendering frequency ambiguity. Thus, we propose two methods to mitigate this problem, which is designed depending on whether the frequency information is available or not.
\subsubsection{Uniform MFC}
\begin{figure}[tbp]
	\setlength{\abovecaptionskip}{-0.1cm}
	\setlength{\belowcaptionskip}{-0.3cm}
	\centerline{\includegraphics[width=0.5\textwidth]{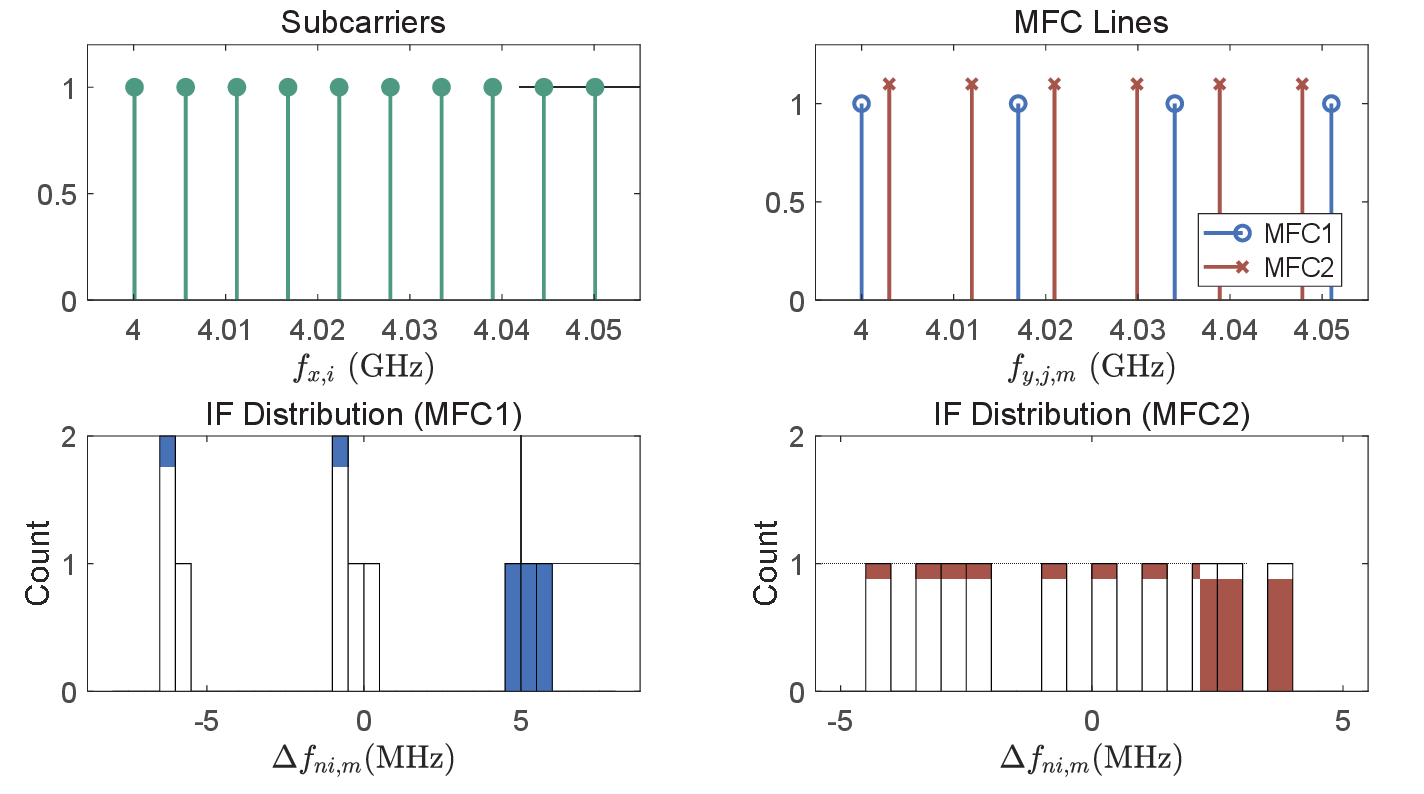}}
		\vspace{0.4em}
	\caption{\textcolor{black}{Illustration of using MFCs to resolve the IF ambiguity. The overlapping IF distributions for MFC1 demonstrate the ambiguity issue, which is mitigated by replacing with MFC2.}}
	\label{comb12}
\end{figure}
For the case where the subcarrier frequencies are unknown, a predefined set of MFC lines with distinct repetitive frequencies can be employed in the MC-RAQR. This configuration is determined offline without increasing system complexity. The MFC lines are designed such that the resulting IF signals are temporally orthogonal. As shown in Fig.~\ref{comb12}, $N=10$ subcarriers occupy a bandwidth of $50$ MHz. After mixing with MFC1, it is obvious that two subcarriers produce the same IF frequency at $\Delta f_{ni,m}=-1.5$ MHz and $\Delta f_{ni,m}=-6.5$ MHz, respectively. By choosing the repetitive frequency of MFC2 properly, all subcarriers can be guaranteed to generate different IF frequencies within $5$ MHz, and thus can be processed separately.

\subsubsection{Non-Uniform MFC}
\begin{figure}[tbp]
	\setlength{\abovecaptionskip}{-0.1cm}
	\setlength{\belowcaptionskip}{-0.3cm}
	\centerline{\includegraphics[width=0.48\textwidth]{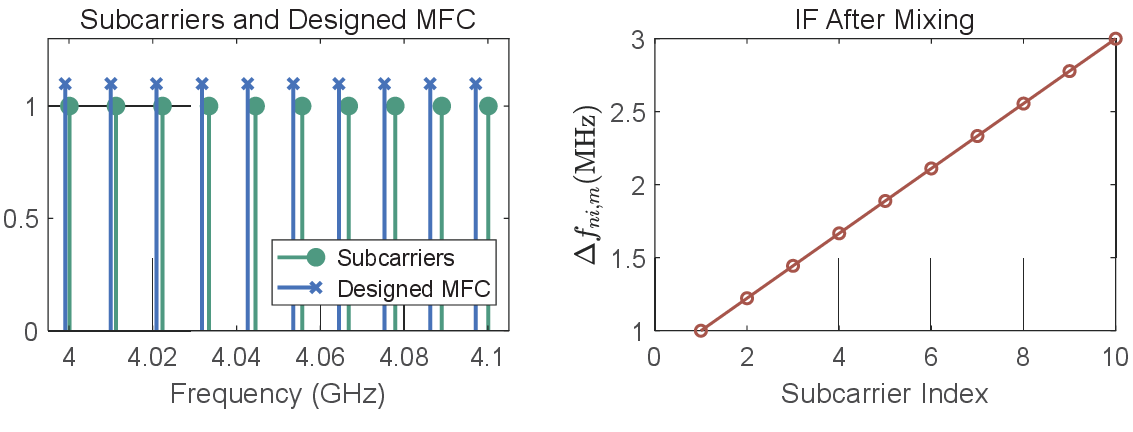}}
		\vspace{0.4em}
	\caption{\textcolor{black}{Illustration of the proposed non-uniform MFC scheme to achieve linear and non-overlapping IF distribution.}}
	\label{nu_MFC}
    \vspace{-0.6em}
\end{figure}
For sensing or wireless communication with prior knowledge of the subcarrier frequencies, we can design non-uniform MFC lines to completely remove frequency ambiguity of IF signal. For instance, in Fig. \ref{nu_MFC}, $N=10$ subcarriers with $100$ MHz bandwidth can be completely separated by letting $\Delta f_{ni,m}=i\delta$, where $\delta$ is the frequency interval between adjacent IF frequencies. By doing so, not only orthogonality between $\Delta f_{ni,m}$ can be guaranteed, but we can also let the range of IF signal strictly within the instantaneous bandwidth.

In practice, the signal impinging at the $m$-th sensor should be the transmit signal propagating through the complex environment with some delay, which is critical for distance and AoA estimation in multi-carrier sensing. In this case, the phase information can be written as 
\begin{equation}
		\setlength\abovedisplayskip{2.7pt}
	\setlength\belowdisplayskip{2.7pt}
\label{phase}
    \Delta {\phi _{ni,m}} = \tilde{\theta}_{i,m}-\varphi_{ni,m}-2\pi f_{x,i}\left(\tau_i+\frac{(m-1)d\sin\theta}{c}\right),
\end{equation}
where $ \tilde{\theta}_{i,m}$ is the phase of the transmit signal. $\tau_i$ is the overall delay of the $i$th subcarrier signal and $\theta\in [-\frac{\pi}{2},\frac{\pi}{2}]$ is the direction of the RF signal with respect to the normal vector of the sensor array.
\begin{theorem}\label{basebandim}
The baseband signal of the $i$-th carrier at the $m$-th sensor used for signal processing can be expressed as
\begin{equation}
	\setlength\abovedisplayskip{2.7pt}
	\setlength\belowdisplayskip{2.7pt}
\label{vmi}
\begin{aligned}
{y_{m,i}}(t) = \kappa_m h_{m,i}{s_{i}}(t),
\end{aligned}
\end{equation}
where  we assume the phase of MFC comb line at the $m$-th sensor remains same for all the $N$ carriers, i.e., ${\varphi _m} = {\varphi _{ni,m}},\forall i$ and $\kappa_m = \rho_m e^{-j\varphi_m}$; $h_{m,i} = \tilde{h}_{m,i}{r_{m,i}}{a_{m}}$ includes the influence of propagation environment with ${r_{m,i}} = {e^{ - j{2\pi }f_{x,i}\tau_i}}$ and ${a_{m}} = {e^{ - j2\pi \frac{f_{x,i}}{c} md\sin \theta }}$; ${s_{i}}(t)$ denotes the transmit signal at the $i$-th subcarrier.
\begin{IEEEproof}
By assuming $N$ subcarriers are temporally orthogonal, the baseband signal of the $i$th carrier in (\ref{vmtilde}) is 
\begin{equation}
		\setlength\abovedisplayskip{2.7pt}
	\setlength\belowdisplayskip{2.7pt}
\label{baseband_i}
    {{\tilde v}_{m,i}}(t) = {\rho _m}\sqrt {{{\cal P}_{x,i,m}}} {e^{j\Delta {\phi _{ni,m}}}}.
\end{equation}
By further substituting (\ref{phase}) into (\ref{baseband_i}), we obtain \eqref{vmi}.
\end{IEEEproof}
\end{theorem}

Theorem \ref{basebandim} shows that the baseband signal to be processed includes the following three parts: 

$\kappa_m$: It is the coefficients related to the configuration of the MC-RAQR, including the Rabi frequency of the probe and coupling laser, the number of comb lines in the MFC field, the power and phase of the local optical signal, and the detuing of the MFC field that enables continuous frequency detection, as shown in (\ref{closed_kappa_m1}) at the top of next page.

${h_{m,i}}$: It includes the complex coefficient related to the environment and the phases related to the time delay during the signal propagation. \textcolor{black}{Note that in the models of previous work, only $a_{m,i}$ is considered for the AoA detection  \cite{doa_robinson,gong_doa}. This is because the delay resolution is inversely proportional to the signal bandwidth according to $\Delta_\tau = 1/B$. For a monostatic sensing setup, this translates to a spatial distance resolution of $\Delta_r = \frac{c}{2B}$ \cite{crb1}, where $c$ is the speed of light. If conventional RAQRs are utilized, their restricted instantaneous bandwidth inherently leads to a coarse distance resolution, making it challenging to distinguish closely spaced targets. By contrast, the bandwidth can be significantly broadened by exploiting the proposed MC-RAQR, which fundamentally refines the distance resolution and enables accurate distance estimation for dense-target sensing.}

${s_{i}}(t)$: It is the baseband signal of the $i$-th carrier at the transmitter. This indicates that the transmit signal may be fully extracted by analyzing $v_{m,i}(t)$, thus enabling the recovery of communication information.

\begin{figure*}[ht] 
	\centering
	\begin{equation}
		\setlength\abovedisplayskip{2.7pt}
		\setlength\belowdisplayskip{2.7pt}
		\begin{aligned}
			{\kappa _m} = \underbrace {\frac{{\eta q\sqrt \pi  }}{{{\hbar ^2}c\sqrt {\ln 2} }}}_{{\rm{Constant}}} \times \underbrace {{F_p}d{\mu _{45}}\left| {{U_{0,m}}} \right|\sqrt {\frac{{{G_{\text{LNA}}}{\mathcal{P}_{l,m}}}}{{{A_e}}}} }_{\scriptstyle{\rm{Probe~laser,~Local~optical}}\hfill\atop
				\scriptstyle{\rm{~signal~sensor~spacing}}\hfill} \times \underbrace {\frac{{1}}{{\sqrt B }}{e^{ - (j\varphi_m+{k_p}L\mathcal{I}[{\chi _m}])}}\sqrt {{\mathcal{I}^2}[{\chi _m}^\prime ] + {\mathcal{R}^2}[{\chi _m}^\prime ]} \cos ({\phi _{l,m}} - {\phi _{p,m}} + {\psi _{p,m}})}_{\scriptstyle{\rm{Incident~RF signals:~Auxiliary~signal,~Comb~signal }}\hfill\atop
				\scriptstyle{\rm{and~Multi - carrier~communication/sensing~signal}}\hfill}.
			\label{closed_kappa_m1}
		\end{aligned}
	\end{equation}	
		\hrulefill
\end{figure*}

\subsection{{Noise Sources in MC-RAQR}}
Note that (\ref{vmi}) is the noiseless baseband signal through MC-RAQR, which suffers from several noise sources in practice, including the quantum projection noise (QPN), the photon shot noise (PSN) and the thermal noise (ITN) \cite{gongmodel}. Specifically, the power of QPN can be expressed as \cite{bbr}
\begin{equation}
		\setlength\abovedisplayskip{2.7pt}
	\setlength\belowdisplayskip{2.7pt}
\sigma _{QPN,i}^2 = {\rho _m}{B_i}\frac{{{\hbar ^2}({\Gamma _{nat}} + {\Gamma _{bbr}})}}{{\mu _{45}^2{N_0}{\Upsilon _1}{\Upsilon _2}}},
\end{equation}
where $B_i$ denotes the bandwidth of the $i$-th carrier, $\Upsilon_1$ and $\Upsilon_2$ denote the population rate and the volume containing atoms, respectively. $\Gamma_{nat}$ and $\Gamma_{bbr}$ represent the natural dephasing rate and black-body radiation induced dephasing rate, respectively. 
Furthermore, the PSN is induced in the transformation from the optical signal to the photocurrent, having a power of \cite{psn}
\begin{equation}
		\setlength\abovedisplayskip{2.7pt}
	\setlength\belowdisplayskip{2.7pt}
	\begin{aligned}
	&\sigma _{PSN,i}^2 = 2q{B_i}\alpha ({{\cal P}_{l,m}} + {{\cal P}_m}({\Omega_{y,m}})).
	\end{aligned}
\end{equation}
Finally, the ITN comes from LNA and other electronic circuits, thus can be modeled as $\sigma _{ITN,i}^2 = {k_B}T{B_i}{G_{\text{LNA}}}$\cite{itn}, where $k_B$ is the Boltzmann constant and $T$ is the temperature. Based on the above-mentioned discussion, the power at the baseband of the MC-RAQR can be expressed as
\begin{equation}
	\setlength\abovedisplayskip{2.7pt}
	\setlength\belowdisplayskip{2.7pt}
	\label{noise}
\sigma _{w,i}^2 = \frac{1}{2} ({\sigma _{QPN,i}^2 + \sigma _{PSN,i}^2 + \sigma _{ITN,i}^2}).
\end{equation}

\section{Performance analysis for MC-RAQR Aided Communication and Sensing}\label{performance_analysis}
The impinging signal $x_{m,i}(t)$ can be the uplink wireless communication signal or an echo signal from the target for sensing system. In the following, we first present the MC-RAQR for communication and then for sensing by detailing their models and performance, respectively. 

\subsection{Wireless Communication Model}
If $s_{i}(t)$ is the uplink signal transmitted from the $i$-th user equipment (UE), the discrete-time baseband signal for MC-RAQR aided wireless communication can be expressed as
\begin{equation}
	\setlength\abovedisplayskip{2.7pt}
	\setlength\belowdisplayskip{2.7pt}
	\label{dt_com}
	{y_{m,i}}[n] = {\kappa _m}{h_{i,m}}{s_i}[n] + w_{i}[n],
\end{equation}
where we assume $w_{m,i}[n]$ follows a complex additive white Gaussian noise distribution, i.e., $w_{m,i}[n] \sim \mathcal{CN}(0,\sigma _{w,i,m}^2)$, and $\sigma _{w,i,m}^2$ is given by (\ref{noise}). By concatenating ${y_{m,i}}[n]$ for all $m$, we obtain the vector form of (\ref{dt_com}) as
\begin{equation}
		\setlength\abovedisplayskip{2.7pt}
	\setlength\belowdisplayskip{2.7pt}
{{\bf{y}}_i} = {\bm K}{{\bf{h}}_i}{s_i} + {{\bf{w}}_i},
\end{equation}
where ${\bm K} ={ \rm {diag}}\{\kappa_m\}$; ${\bf {h}}_i = [h_{i,1},...,h_{i,M}]^T$ denotes the channel vector of the $i$-th carrier. \textcolor{black}{Upon employing the well-known maximum ratio combining vector ${\bf{f}}_i = \frac{{{{ {{\bm K}{\bf{}}{{\bf{h}}_i}} }}}}{{\left\| {{\bm K}{\bf{}}{{\bf{h}}_i}} \right\|}}$ \cite{MRC}, we can formulate the output signal as
	\begin{equation}
		\label{MRC}
		\setlength\abovedisplayskip{2.7pt}
		\setlength\belowdisplayskip{2.7pt}
		\begin{aligned}
			{y_i} = {\bf{f}}_i^H{{\bf{y}}_i} = \left\| {{\bm K}{\bf{}}{{\bf{h}}_i}} \right\|{s_i} + \frac{{{{\left( {{\bm K}{\bf{}}{{\bf{h}}_i}} \right)}^H}}}{{\left\| {{\bm K}{\bf{}}{{\bf{h}}_i}} \right\|}}{{\bf{w}}_i}.
		\end{aligned}
	\end{equation}
	It is worth noting that the combining scheme in (\ref{MRC}) assumes the availability of perfect channel state information (CSI) at the receiver, which is widely adopted in wireless communication literature. Since the proposed MC-RAQR model encapsulates quantum-optical parameters within a classical framework, conventional pilot-based channel estimation methods may still be applied in practical deployments \cite{CSI1,CSI2}.} Therefore, the received SNR for the $i$-th carrier can be obtained as
\begin{equation}
	\setlength\abovedisplayskip{2.7pt}
	\setlength\belowdisplayskip{2.7pt}
	\label{snr1}
{\gamma _i} = \frac{{{\mathcal{P}_c}{{\left\| {{\bm K}{{\bf{h}}_i}} \right\|}^2}}}{{\sigma _{w,i}^2}} = \frac{{2{\mathcal{P}_c}{{\left\| {{\bm K}{{\bf{h}}_i}} \right\|}^2}}}{{\sigma _{QPN,i}^2 + \sigma _{PSN,i}^2 + \sigma _{ITN,i}^2}},
\end{equation}
where $\mathcal{P}_c$ denotes the power of the transmit baseband signal $s_i(t)$. Assume channel matrix can be written as ${h_{i,m}} = \frac{{\sqrt {{\beta _i}} }}{r}{e^{ - j\frac{{2\pi }}{c}\Delta {f_{ni,m}}(r + md\sin \theta )}}$, where $\beta_i$ denotes the complex-valued channel gain, and $r$ being the distance between the UE and MC-RAQR, then (\ref{snr1}) can be expressed as
\begin{equation}
	\setlength\abovedisplayskip{2.7pt}
	\setlength\belowdisplayskip{2.7pt}
{\gamma _i} = \frac{{{2{\bar {\mathcal{P}}}_{c,i}}\sum\limits_{m = 1}^M {{\left| {{\kappa _m}} \right|^2}} }}{{\sigma _{QPN,i}^2 + \sigma _{PSN,i}^2 + \sigma _{ITN,i}^2}},
\end{equation}
where ${{\bar {\mathcal{P}}_{c,i}}} = \frac{{{\mathcal{P}_c}{\beta _i}}}{r^2}$ represents the received power. Collecting all the $N$ carriers, we obtain the channel capacity as follows 
\begin{equation}
		\setlength\abovedisplayskip{2.7pt}
	\setlength\belowdisplayskip{2.7pt}
\begin{aligned}
    R &= {\sum\limits_{i = 0}^{N-1} {{B_i}{{\log }_2}(1 + {\gamma _i})} }.\\
\end{aligned}
\end{equation}
\subsection{Wireless Sensing Model}
Once the impinging signals are echoes reflected from $K$ targets, the discrete-time baseband signal for sensing becomes
\begin{equation}
		\setlength\abovedisplayskip{2.7pt}
	\setlength\belowdisplayskip{2.7pt}
	\label{sensing_1}
	{y_{m,i}}[n] = \kappa_m\sum\limits_{k = 1}^K {{\alpha _k}{r_{m,i}}(r_k){a_{m,i}}(\theta_k )s[n]}  + {w_i}[n],
\end{equation}
where $\alpha_k$ denotes the reflection coefficient of target $k$, which includes the effect of the radar cross section (RCS), and the transmit beamforming gain; $s[n]$ is the transmitted sensing waveform, which remains the same across different carrier frequencies. Note that in this paper, we consider the monostatic sensing with a double-hop transmission, which means $\tau_k=\frac{2r_k}{c}$. Furthermore, we assume the frequencies of subcarriers satisfy $f_{x,i} = f_c+i\Delta f$, where $f_c$ is the carrier frequency and $\Delta f$ is the interval of adjacent subcarrier frequencies. Based on the above discussions, we reformulate (\ref{sensing_1}) as
\begin{equation}
	\setlength\abovedisplayskip{2.7pt}
	\setlength\belowdisplayskip{2.7pt}
\label{sensing_2}
{{\bf{y}}_i}[n] = {\kappa _m}\sum\limits_{k = 1}^K {{{\tilde \alpha }_k}{e^{ - j\frac{{4\pi }}{c}i\Delta f{r_k}}}{\bf{a}}({\theta _k})s[n]}  + {{\bf{w}}_i}[n],
\end{equation}
where $\tilde{\alpha}_k = \alpha_k{e^{ - j\frac{4\pi }{c}f_c r_k}}$; ${[{\bf{a}}(\theta )]_m} = {e^{ - j\frac{{2\pi }}{{{\lambda _c}}}md\sin \theta }}$ is the steering vector and $\lambda_c$ denotes the wavelength of the carrier frequency. It can be observed that the unknown parameters, namely distance $r_k$ and AoA $\theta_k$ for each target $k$, are not coupled in the phase of ${{\bf{y}}_i}[n]$, so they can be estimated independently. In the following, we first study how to estimate the AoA, as its resolution is inversely proportional to the number of sensors, which can be significantly higher than the resolution of distance estimation. 
\subsubsection{AoA Estimation} 
For multi-carrier AoA detection, the information from different subcarriers can be regarded as spatial snapshots, enabling array signal processing techniques such as the MUSIC algorithm to be applied. In particular, for each carrier frequency $i$, we can reformulate (\ref{sensing_2}) as 
\begin{equation}
		\setlength\abovedisplayskip{2.7pt}
	\setlength\belowdisplayskip{2.7pt}
	\label{aoa1}
{{\bf{y}}_i}[n] = {\bm{K }}{{\bf{A}}}{{\bf{s}}_{1,i}}[n] + {{\bf{w}}_i}[n],
\end{equation}
where ${{\bf{A}}} = [{{\bf{a}}}({\theta _1}),...,{{\bf{a}}}({\theta _K})] \in {\mathbb{C}^{M \times K}}$ denotes the steering matrix for $K$ targets. ${{\bf{s}}_{1,i}}[n] = {[{\tilde{\alpha} _1}{e^{ - j\frac{{4\pi }}{c}{i\Delta f}{r_1}}},...,{\tilde{\alpha} _K}{e^{ - j\frac{{4\pi }}{c}{i\Delta f}{r_K}}}]^T}s[n] \in {\mathbb{C}^{K \times 1}}$ represents the equivalent baseband signal. By accumulating $J$ snapshots, the matrix form of (\ref{aoa1}) can be expressed as
\begin{equation}
		\setlength\abovedisplayskip{2.7pt}
	\setlength\belowdisplayskip{2.7pt}
	\label{matrix_aoa}
	{{\bf{Y}}_i} = {\bm{K}}{{\bf{A}}}{{\bf{S}}_1} + {{\bf{W}}_i} \in {\mathbb{C}^{M \times J}},
\end{equation}
where ${{\bf{Y}}_i}=[{{\bf{y}}_i}[1],...,{{\bf{y}}_i}[J]]$. Furthermore, we concatenate ${{\bf{Y}}_i}$ across all subcarriers to construct virtual array snapshots, and compute its sample covariance matrix as
\begin{equation}
		\setlength\abovedisplayskip{2.7pt}
	\setlength\belowdisplayskip{2.7pt}
	{{\bf{R}}} = \frac{1}{NJ}{{\bf{Y}}}{\bf{Y}}^H= {{\bf{U}}_{s}}{{\bf{\Sigma }}_{s}}{\bf{U}}_{s}^H + {{\bf{U}}_{n}}{{\bf{\Sigma }}_{n}}{\bf{U}}_{s}^H,
\end{equation}
where ${{\bf{Y}}} = [{{\bf{Y}}}_1,...,{{\bf{Y}}}_N]$. ${{\bf{\Sigma }}_s}$ and ${{\bf{\Sigma }}_n}$ are diagonal matrices with the $K$ largest eigenvalues and the remaining $M-K$ small eigenvalues as their diagonal elements, respectively. ${{\bf{U}}_s}$ and ${{\bf{U}}_n}$ denote the eigenvectors that span the signal and noise subspaces, respectively, which can be obtained by performing eigenvalue decomposition (EVD) of ${{\bf{R}}}$. Therefore, the MUSIC spectrum for AoA estimation can be written as
\begin{equation}
	\setlength\abovedisplayskip{2.7pt}
	\setlength\belowdisplayskip{2.7pt}
	{P}(\theta ) = \frac{1}{{{\bf{a}}^H(\theta ){{\bm{K}}^H}{{\bf{U}}_{n}}{\bf{U}}_{n}^H{\bm{K}}{{\bf{a}}}(\theta )}}.
		\label{aoaP}
\end{equation}
The estimated value of AoAs, denoted as $\hat{\theta}_k$, can be obtained by finding the $K$ largest peaks of (\ref{aoaP}).

\subsubsection{Distance Estimation}
After obtaining the $K$ estimated AoAs, spatial domain beamforming can be performed for harnessing the beamforming gain and to decouple the signals in the spatial domain. In particular, for the $k$th target, upon exploiting the beamforming vector ${\bf{b}}({\hat \theta _k}) = {{\bm {K} \bf a}}({\hat \theta _k})$ to (\ref{sensing_2}), we have  
\begin{equation}
	\setlength\abovedisplayskip{2.7pt}
	\setlength\belowdisplayskip{2.7pt}
	\label{bf}
	\begin{aligned}
	{y_i}[n] &= \frac{{{{\bf{b}}^H}({{\hat \theta }_k})}}{{\left\| {{{\bf{b}}^H}({{\hat \theta }_k})} \right\|}}{{\bf{y}}_i}[n]\\
	&= \left\| {{{\bf{b}}^H}({{\hat \theta }_k})} \right\|{\tilde \alpha _k}{e^{ - j\frac{{4\pi }}{c}i \Delta f{r_k}}}s[n] + {\bar {w}_i}[n],
	\end{aligned}
\end{equation}
where ${\bar {w}_i}[n] = \left\| {{{\bf{b}}^H}({{\hat \theta }_k})} \right\| w_i[n]$. By further considering information on all the subcarriers and snapshots, the matrix form of (\ref{bf}) becomes 
\begin{equation}
	\setlength\abovedisplayskip{2.7pt}
	\setlength\belowdisplayskip{2.7pt}
	\label{matrix_r}
{{\bf{Y}}_k} = {{\bar \alpha }_k}{{{\bf{c}}(r)}\bf{s}} + { {\bf{\bar W}}_k} \in{\mathbb{C}^{N \times J}},
\end{equation}
where ${{\bar \alpha }_k} = \| {{{\bf{b}}^H}({{\hat \theta }_k})} \|{\tilde \alpha _k}$ represents the beamforming gain and reflection coefficient of target $k$; ${[{{\bf{c}}}(r)]_i} = {e^{ - j\frac{{4\pi }}{c}i \Delta fr}}$ and ${\bf{s}} = [s[1],...,s[J]]$. Similar to the procedures in AoA estimation, by calculating the covariance matrix of (\ref{matrix_r}) as ${{\bf{R}}_k} = \frac{1}{J}{{\bf{Y}}_k}{\bf{Y}}_k^H$ and performing the EVD, we can obtain the corresponding noise subspace ${{{\bf{U}}_n}}$. For each target $k$ with estimated AoA $\hat{\theta}_k$, we construct the MUSIC spectrum of the distance estimation as
\begin{equation}
	\setlength\abovedisplayskip{2.7pt}
	\setlength\belowdisplayskip{2.7pt}
	{P_{k}}(r) = \frac{1}{{{\bf{c}}^H(r){{\bf{U}}_{n}}{\bf{U}}_{n}^H{{\bf{c}}}(r)}}.
\end{equation}
The estimated value of distance for target $k$ can be acquired by targeting the largest peak of ${P_{k}}(r)$.

\subsubsection{Cram\'er-Rao Bound (CRB) Analysis}
To further evaluate the estimation performance of the above parameters, we employ the CRB, serving as a lower bound for characterizing any unbiased mean-square error (MSE) estimator. To begin with, we express (\ref{sensing_2}) in vector form as
\begin{equation}
	\setlength\abovedisplayskip{2.7pt}
	\setlength\belowdisplayskip{2.7pt}
	{{\bf{y}}_i}[n] = \sum\limits_{k = 1}^K {{\tilde \alpha _k}{{\bf{h}}_{s,i}}({r_k},{\theta _k})s[n]}  + {{\bf{w}}_i}[n],
\end{equation}
where ${[{{\bf{h}}_{s,i}}({r_k},{\theta _k})]_m} = {\kappa _m}{e^{ - j\frac{4\pi }{c}i\Delta f r_k}} {e^{ - j \frac{2\pi}{\lambda_c} md\sin \theta_k }}$. Assume all $K$ targets can be well-separated and let ${\bf{z}} = {[{\theta _k},{r_k}]^T}$ include those unknown parameters of target $k$. Upon defining ${{\bf{g}}_i}[n] \triangleq {\tilde \alpha _k}{{\bf{h}}_{s,i}}({r_k},{\theta _k})s[n]$, we formulate the Fisher information matrix (FIM) of the $i$-th carrier with respect to $\bf z$ as \cite{crb1}
\begin{equation}
	\setlength\abovedisplayskip{2.7pt}
	\setlength\belowdisplayskip{2.7pt}
	\small
	\label{fi}
	\begin{aligned}
		{{\bf{F}}_i} &= \frac{2}{{{\sigma _{w,i}^2}}}\sum\limits_{n = 1}^J {{\cal R}\left\{ {\frac{{\partial {{\bf{g}}_i}[n]}}{{\partial {\bf{z}}}}{{\left( {\frac{{\partial {{\bf{g}}_i}[n]}}{{\partial {\bf{z}}}}} \right)}^H}} \right\}} \\
		&= \frac{2}{{{\sigma _{w,i}^2}}}\left[ {\begin{array}{*{20}{c}}
				{\sum\limits_{n = 1}^J {{{\left\| {\frac{{\partial {{\bf{g}}_i}[n]}}{{\partial {\theta _k}}}} \right\|}^2}} }&{\sum\limits_{n = 1}^J {\frac{{\partial {{\bf{g}}_i}[n]}}{{\partial {\theta _k}}}{{\left( {\frac{{\partial {{\bf{g}}_i}[n]}}{{\partial {r_k}}}} \right)}^H}} }\\
				{\sum\limits_{n = 1}^J {\frac{{\partial {{\bf{g}}_i}[n]}}{{\partial {\theta _k}}}{{\left( {\frac{{\partial {{\bf{g}}_i}[n]}}{{\partial {r_k}}}} \right)}^H}} }&{\sum\limits_{n = 1}^J {{{\left\| {\frac{{\partial {{\bf{g}}_i}[n]}}{{\partial {r_k}}}} \right\|}^2}} }
		\end{array}} \right].
	\end{aligned}
\end{equation}
By furthermore combining all $N$ carriers, we obtain the total FIM as ${\bf{F}} = \sum_{i = 1}^N {{{\bf{F}}_i}}$. 

\begin{theorem}\label{crb}
	For monostatic MC-RAQR sensing, the CRBs of estimating the AoA $\theta_k$ and distance $r_k$ are derived as 
	\begin{equation}
    \small
    \setlength\abovedisplayskip{2.7pt}
		\setlength\belowdisplayskip{2.7pt}
		\label{crbrt}
		\begin{aligned}
				\mathrm{CRB}({\theta _k}) &= \frac{{\lambda _c^2}}{{4{\pi ^2}J{{\overline {\mathcal{P}} }_{i,k}}{d^2}{{\cos }^2}{\theta _k}}}\\
				&\times \frac{{2N + 1}}{{N\sum\limits_{m = 1}^M {{{\left| {{\kappa _m}} \right|}^2}} ((4N + 2){\mu _2} - 3(N + 1)\mu _1^2)}},\\
			\mathrm{CRB}({r_k}) &= \frac{{{c^2}}}{{32{\pi ^2}J{{\overline {\mathcal{P}} }_{i,k}}\Delta {f^2}}}\\
			&\times \frac{{{\mu _2}}}{{{N}(N + 1)\sum\limits_{m = 1}^M {{{\left| {{\kappa _m}} \right|}^2}} \left( {\frac{{(2N + 1)}}{6}{\mu _2} - \frac{{N + 1}}{4}\mu _1^2} \right)}},
		\end{aligned}
	\end{equation}
	where ${{\bar {\mathcal{P}}}_{i,k}} = \frac{{{{\left| \tilde {\alpha}_k  \right|}^2}{{\cal P}_s}}}{{{\sigma _{w,i}^2}}}$ denotes the received SNR at the MC-RAQR; ${\mu _1} = \frac{{\sum_{m = 1}^M {{{\left| {{\kappa _m}} \right|}^2}} m}}{{\sum_{m = 1}^M {{{\left| {{\kappa _m}} \right|}^2}} }}$ and ${\mu _2} = \frac{{\sum_{m = 1}^M {{{\left| {{\kappa _m}} \right|}^2}} {m^2}}}{{\sum_{m = 1}^M {{{\left| {{\kappa _m}} \right|}^2}} }}$ represent the weighted mean and power mean of the index variable $m$ with respect to the weight ${{{\left| {{\kappa _m}} \right|}^2}}$, respectively.
	\begin{IEEEproof}
		Please refer to Appendix D.
	\end{IEEEproof}
\end{theorem}
Theorem \ref{crb} shows that the CRBs of both the AoA and distance decrease as the increase of the MC-RAQR gain $\kappa_m$. Furthermore, when the number of subcarrier $N$ increase, both of the CRBs will decrease, which indicates that our proposed MC-RAQR can significantly improve the estimation accuracy compared with the conventional RAQRs that can only receive single-carrier signal with limited bandwidth.

\begin{corollary}\label{crbcoro}
When the RAQR gains are identical for all $M$ sensors, namely $\kappa_m=\kappa, \forall m,$ the CRBs of both the AoA and distance in \eqref{crbrt} become
\begin{equation}
		\setlength\abovedisplayskip{2.7pt}
	\setlength\belowdisplayskip{2.7pt}
\begin{aligned}
	\mathrm{CRB}({\theta _k}) &= \frac{{3\lambda _c^2}}{{{\pi ^2}J{{\overline {\mathcal{P}} }_{i,k}}{d^2}{{\cos }^2}{\theta _k}}}\\
	&\times \frac{{2N + 1}}{{NM{{\left| \kappa  \right|}^2}(M + 1)(7MN - M - N - 5)}},\\
	\mathrm{CRB}({r_k}) &= \frac{{{c^2}}}{{4{\pi ^2}J{{\overline {\mathcal{P}} }_{i,k}}\Delta {f^2}}}\\
	&\times \frac{{3(2M + 1)}}{{{N}(N + 1)M{{\left| \kappa  \right|}^2}(7MN - M - N - 5)}}.
\end{aligned}
\label{CRB}
\end{equation}
\begin{IEEEproof}
Since $\kappa_m=\kappa$, $\mu_1$ and $\mu_2$ in (\ref{crbrt}) reduce to ${\mu _1} = \frac{{\sum_{m = 1}^M m }}{M} = \frac{{M + 1}}{2}$ and ${\mu _2} = \frac{{\sum_{m = 1}^M {{m^2}} }}{M} = \frac{{(M + 1)(2M + 1)}}{6}$, respectively. Substituting these results of $\mu_1$ and $\mu_2$ into (\ref{crbrt}), we obtain \eqref{CRB} of Corollary \ref{crbcoro}. 
\end{IEEEproof}
\end{corollary}
Compared with those CRBs derived in \cite{crb2}, our CRBs in estimating the AoA and distance in Corollary \ref{crbcoro} become $\frac{1}{{\left| \kappa  \right|}^2}$ smaller than the counterparts of conventional mono-static phased-array radars, rendering a higher estimation accuracy of the proposed MC-RAQR.

\section{Simulation results}\label{simulation}
In this section, we present our simulation results to verify the effectiveness of the proposed MC-RAQR.

\subsection{Experimental Setup}
\textcolor{black}{The ground state and excited state are chosen as $|1\rangle=6S_{1/2}$ and $|2\rangle=6P_{3/2}$, respectively.}
The Rydberg states are chosen as $|3\rangle=60D_{1/2}$, $|4\rangle=60P_{3/2}$ and $|5\rangle=59S_{5/2}$, respectively. The corresponding transition frequency and dipole moment between $|3\rangle$ and $|4\rangle$ are calculated as $f_{34} = 18.177$ GHz and $\mu_{34} = 2057.6\rm{ea_0}$, respectively. Similarly, the transition frequency and dipole moment between $|4\rangle$ and $|5\rangle$ are calculated as $f_{45} = 3.39$ GHz and $\mu_{45} = 3003.34\rm{ea_0}$, respectively. Unless otherwise stated, the sensor length and atomic density are set as $L=10$ cm and $N_0=4.89\times10^{10}$ $\rm{cm^{-3}}$, respectively. Both the probe and coupling lasers are vertically polarized. The probe laser has a beam diameter of $1.7$ mm and a power of $3.8\mu$ W, yielding $\Omega_p=10$ MHz; The coupling laser has the same beam diameter as the probe laser and has a power of $17$mW, yielding $\Omega_c=5.04$ MHz. The decay rate of $|2\rangle$ is set to be $5.2$ MHz. The PD's LNA gain is set to be ${G_{\text{LNA}}}=30$ dB. Without loss of generality, we set the phases of input probe laser and the local optical signal are all zero, i.e., $\phi_{0,m}=\phi_{l,m}=0$. We also assume the RAQR gains are identical for all sensors, i.e., $\kappa_m = \kappa$. Unless otherwise stated, we assume $f_{x,i}=f_c+i\delta_f$, where $f_c = 3.4$ GHz. We assume the QPN noise can be eliminated, and only consider the PSN and ITN noises.

\begin{figure}[tbp]
	\setlength{\abovecaptionskip}{0.1cm}
	\setlength{\belowcaptionskip}{-0.1cm}
	\hspace{-1.1cm}
	\includegraphics[width=0.56\textwidth]{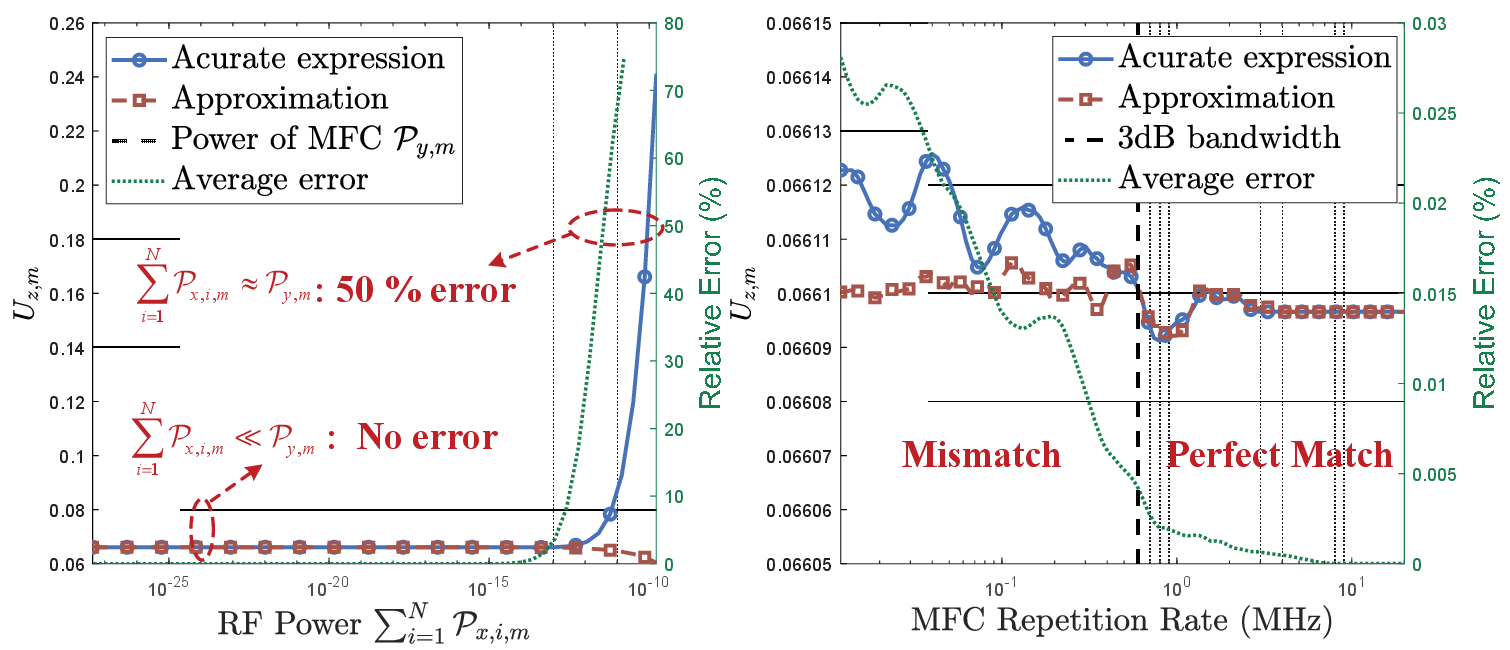}
	\caption{\textcolor{black}{Validation of the accuracy of the proposed model under varying RF power and MFC repetition rate.}}
	\label{error}
\end{figure}
\begin{figure}[tbp]
	\setlength{\abovecaptionskip}{-0.1cm}
	\setlength{\belowcaptionskip}{-0.1cm}
	\centerline{\includegraphics[width=0.45\textwidth]{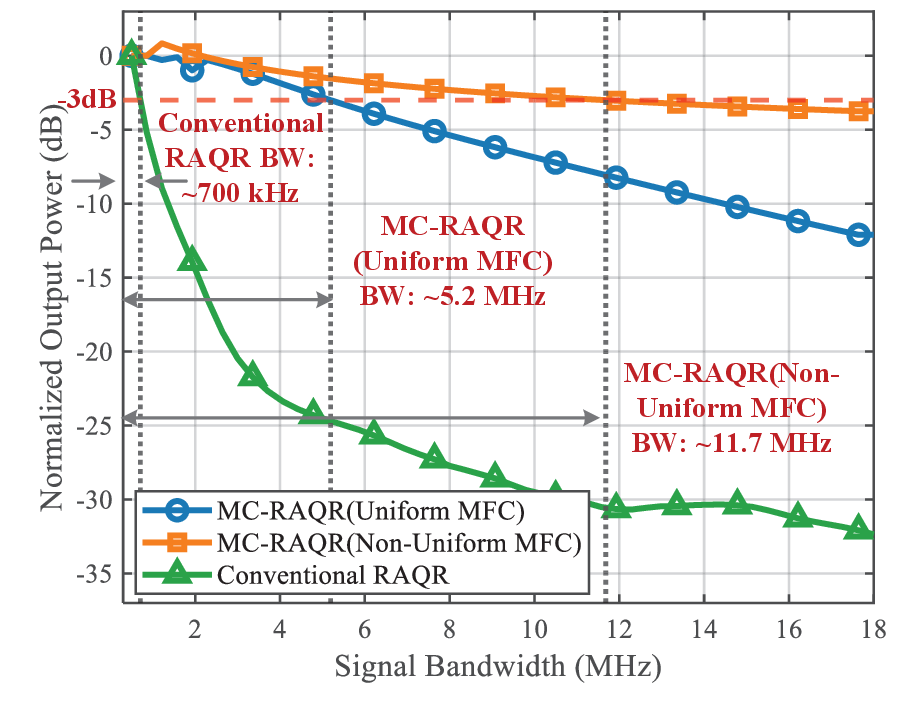}}
	\vspace{0.4em}
	\caption{\textcolor{black}{Comparison of the 3-dB bandwidth among different receiver architectures.}}
	\label{3dbw}
    \vspace{-0.6em}
\end{figure}
\begin{figure*}[t]
		\subfloat{
			\includegraphics[width=0.24\textwidth]{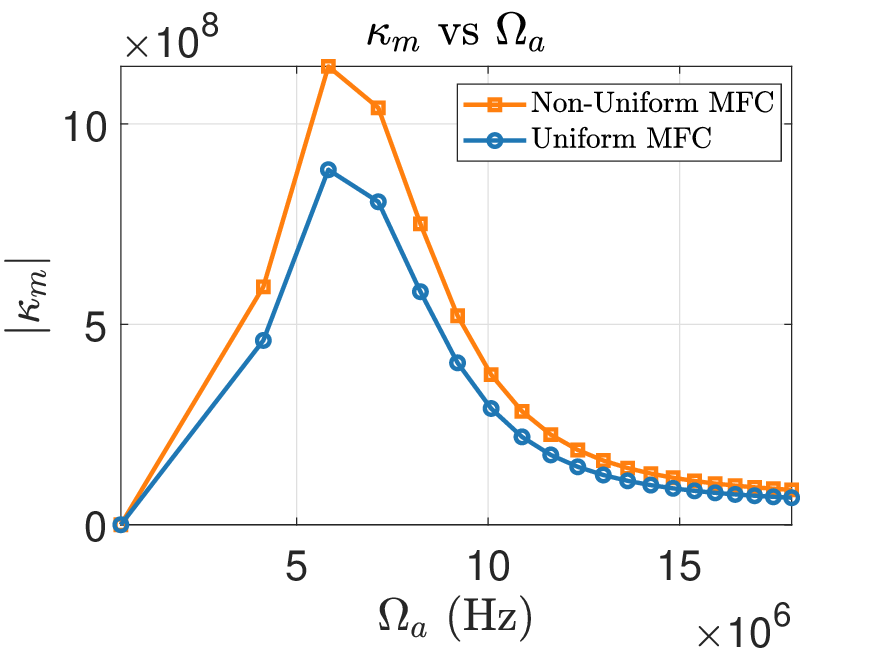}
		}
		\hspace{-4.15ex}
		\subfloat{
			\includegraphics[width=0.24\textwidth]{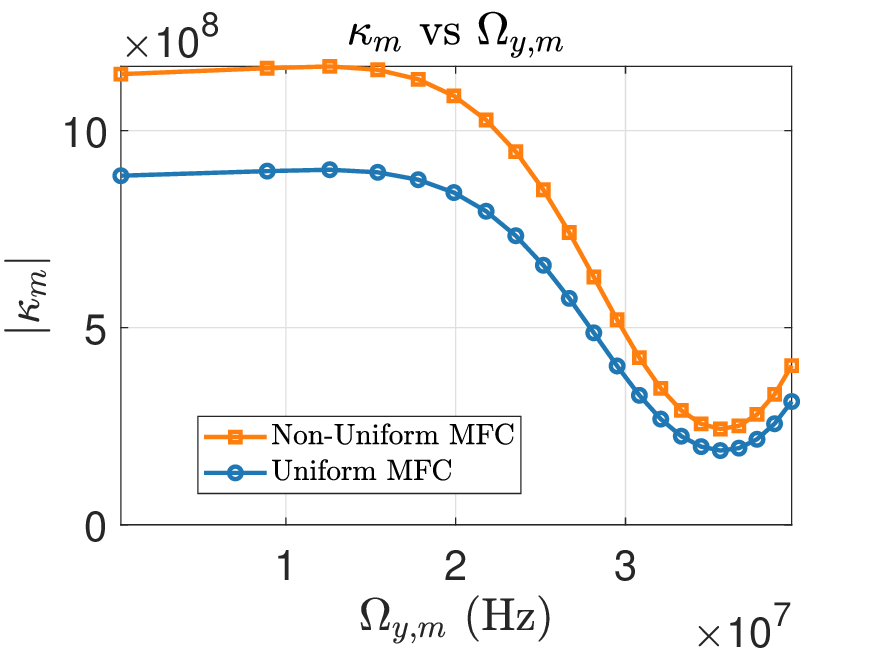}
		}
		\hspace{-4.15ex}
		\subfloat{
			\includegraphics[width=0.24\textwidth]{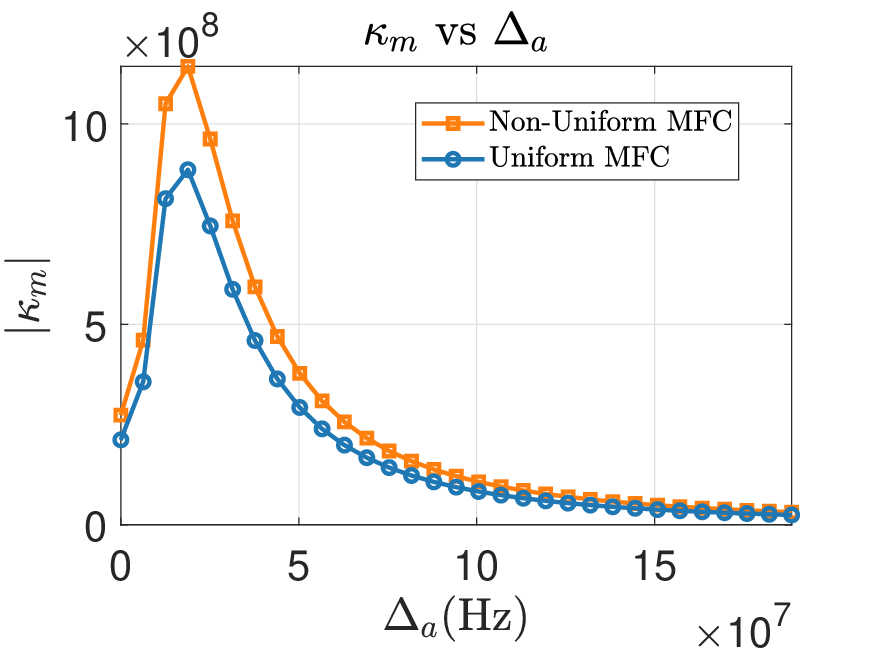}
		}
		\hspace{-4.15ex}
		\subfloat{
			\includegraphics[width=0.24\textwidth]{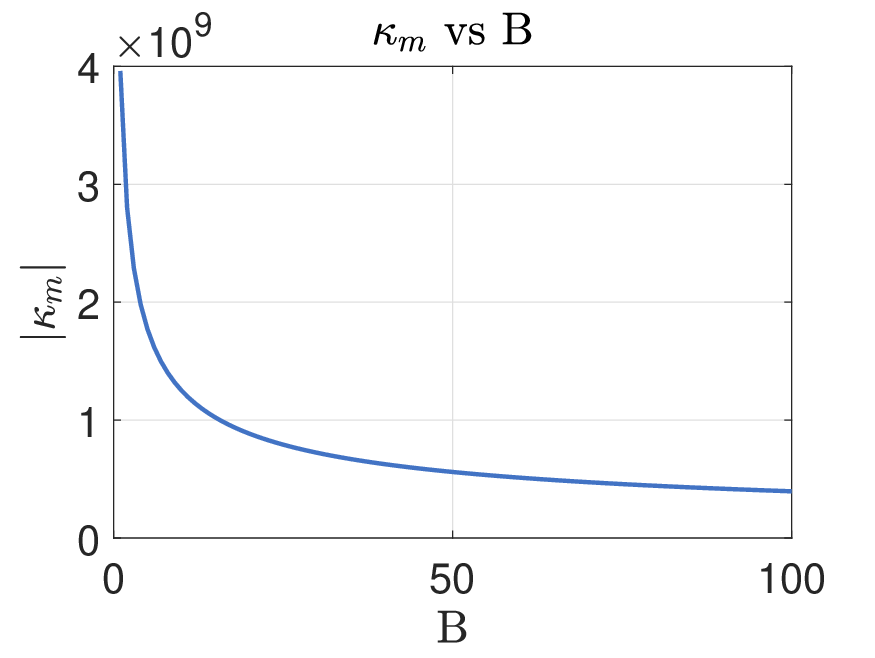}
	}
    \vspace{-0.6em}
	\caption{\textcolor{black}{Evaluation of the MC-RAQR gain $|\kappa_m|$ versus various key optoelectronic parameters under large detuning regime, including the AUX Rabi frequency $\Omega_a$, MFC Rabi frequency $\Omega_{y,m}$, detuning $\Delta_a$, and the number of combs $B$.}}
	\label{ld}
\end{figure*}
\begin{figure*}[t]
	\subfloat{
		\includegraphics[width=0.24\textwidth]{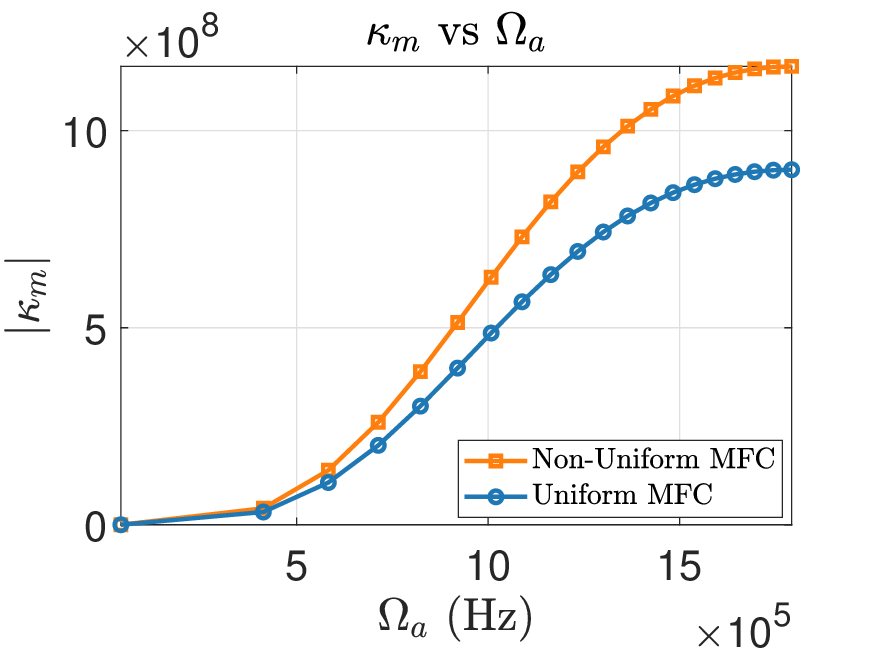}
	}
		\hspace{-4.15ex}
	\subfloat{
		\includegraphics[width=0.24\textwidth]{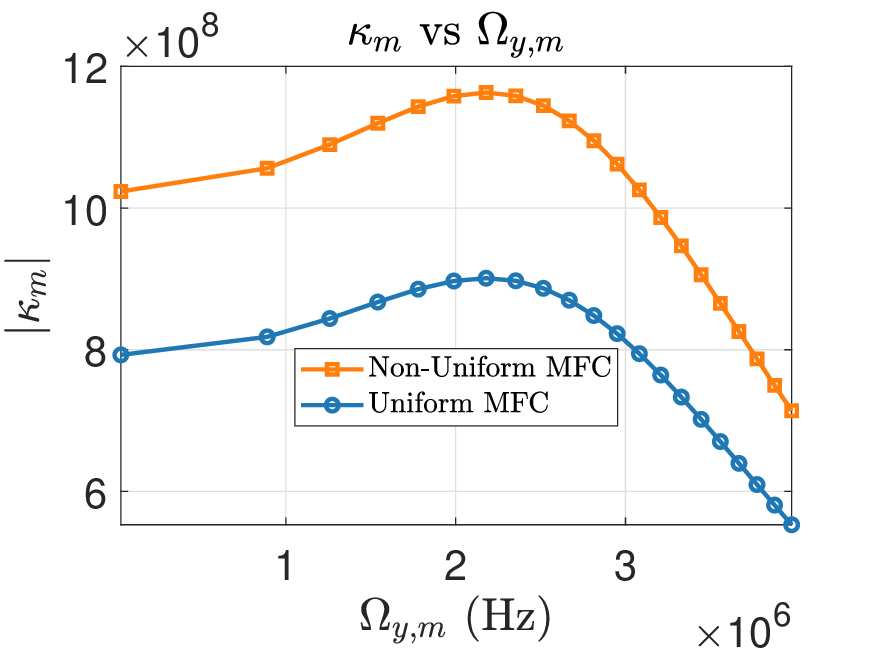}
	}
		\hspace{-4.15ex}
	\subfloat{
		\includegraphics[width=0.24\textwidth]{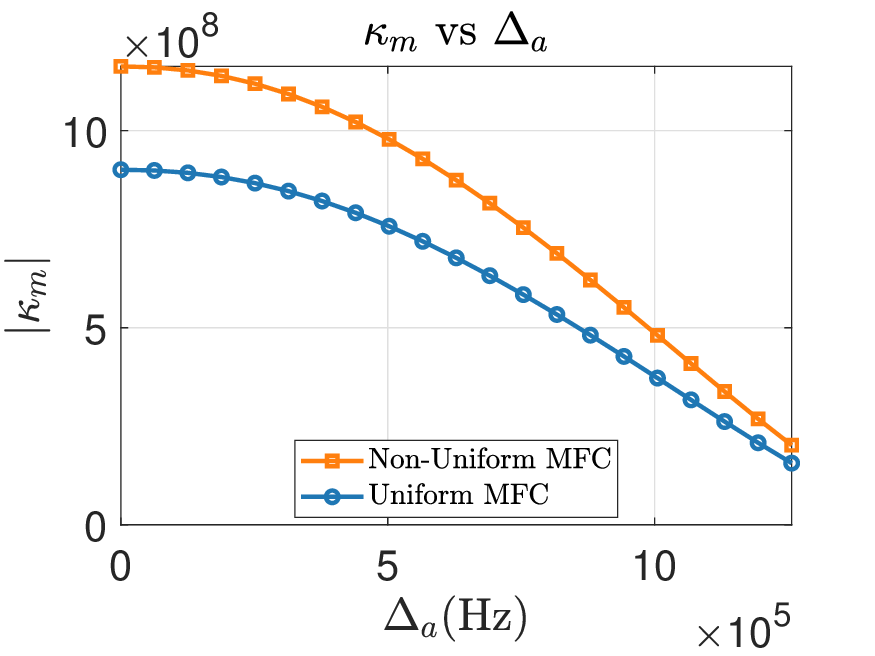}
	}
			\hspace{-4.15ex}
	\subfloat{
		\includegraphics[width=0.24\textwidth]{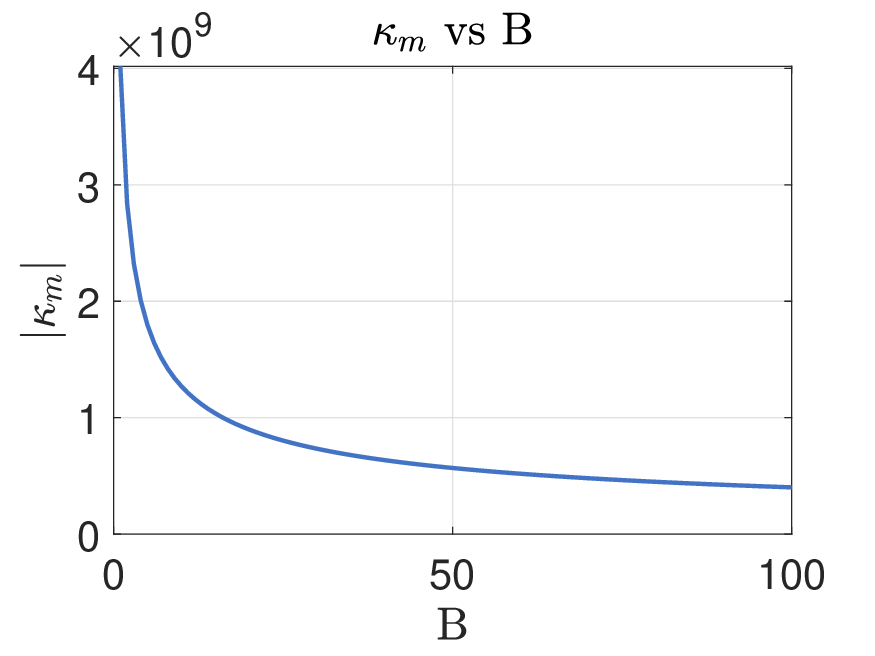}
	}
    \vspace{-0.6em}
	\caption{\textcolor{black}{Evaluation of the MC-RAQR gain $|\kappa_m|$ versus the optoelectronic parameters under small detuning regimes.}}
	\label{sd}
    \vspace{-0.6em}
\end{figure*}

\subsection{Validation of the Proposed Signal Model}
We validate the model of the superimposed signal of MFC and multi-carrier RF signal in Lemma \ref{Uzm}. We set the number of subcarriers to be $N=10$. Fig. \ref{error} (a) shows the result of first-order Taylor approximation in Appendix A error increases as the power of multi-carrier RF signal increases. In particular, when $\sum\nolimits_{i = 1}^N {{P_{x,i,m}}}  \ll {P_{y,m}},$ the error is zero while it can become up to $50\%$ when $\sum\nolimits_{i = 1}^N {{{\cal P}_{x,i,m}}}  \approx {{\cal P}_{y,m}}$. 

Fig. \ref{error} (b) shows the error of approximation when the repetition rate of MFC varies. The 3dB bandwidth is also plotted. It is observed that when the repetition rate is smaller than the 3dB bandwidth, the assumption that ``only the beat signal between each subcarrier and its nearest MFC line is considered" no longer holds, rendering the approximation results can not perfect match the accurate signal. When the repetition rate is larger than the 3dB bandwidth, the approximation holds and the error becomes zero. Note that we consider the uniform MFC in this simulation, for non-uniform MFC, the minimum separation between two MFC lines should be larger than the 3dB bandwidth to make sure the accuracy of our approximation.
\subsection{Parameter Evaluation of MC-RAQR}
Fig. \ref{3dbw} shows the normalized power of extracted signal in (\ref{vmtilde}) versus the signal bandwidth by measuring the output over $10$ symbols. In particular, we utilize 64-order quadrature amplitude modulation ($64$-QAM) modulation for $N=10$ subcarriers signal. We compare the proposed MC-RAQR schemes by exploiting the ``Uniform MFC", and ``Non-Uniform MFC" to the conventional RAQRs to receive the multi-carrier signal. It can be observed that the power of $\tilde{v}_m$ decreases monotonically for all these RAQR schemes, which is reasonable because of the time limit that atoms need to reach the steady state \cite{fall_time}. For larger signal bandwidths, the response time becomes shorter, rendering the atoms difficult to reach the steady state, thus reducing the sensitivity of RAQRs. Compared to the ``conventional RAQR" scheme, employing the MFC can significantly increase the bandwidth of signal that conventional RAQRs can support. Furthermore, the ``Non-Uniform MFC" strategy can support more than $11.7$ MHz bandwidth, which is $17$-fold larger than the ``conventional RAQR" scheme. It is noteworthy that the achievable bandwidth can be further enlarged by designing algorithms to choose $B$ and $\Delta f_{ni,m}$ more wisely.

Fig. \ref{ld} and \ref{sd} show the varying laws of $\kappa_m$ for large and small detuning $\Delta_a$, respectively. Both the ``Uniform MFC" and ``Non-Uniform MFC" are considered for comparison. On one hand, it is shown that for large detuning $\sim 50$ MHz in Fig. \ref{ld}, $|\kappa|$ reaches its peak value at $\Delta_a \approx 25$ MHz and decrease rapidly for larger detuning. It is also observed that $|\kappa|$ reaches the maximal value at $\Omega_{\rm{a}} \approx 7$ MHz, and decreases for both smaller and larger $\Omega_{\rm{a}}$. Furthermore, $|\kappa|$ remains a constant for small $\Omega_{y,m}$ and drops significantly when $\Omega_{y,m}$ increases. On the other hand, for small detuning in Fig.~\ref{sd}, $\kappa$ gradually decreases as $\Delta_{\rm{a}}$ increases. Meanwhile, it becomes monotonically increase with $\Omega_{{\rm{a}}}$ and reaches the maximal point when $\Omega_{y,m} \approx 2$ MHz. For both large and small detuning cases, $|\kappa|$ monotonically decreases as the number of MFC lines grows. Therefore, the ``Non-Uniform MFC" scheme always outperforms the ``Uniform MFC" scheme since smaller $B$ can be designed.

%

\subsection{Performance Analysis of Communication and Sensing}
\begin{figure}[tbp]
	\setlength{\abovecaptionskip}{-0.1cm}
	\setlength{\belowcaptionskip}{-0.1cm}
	\centerline{\includegraphics[width=0.45\textwidth]{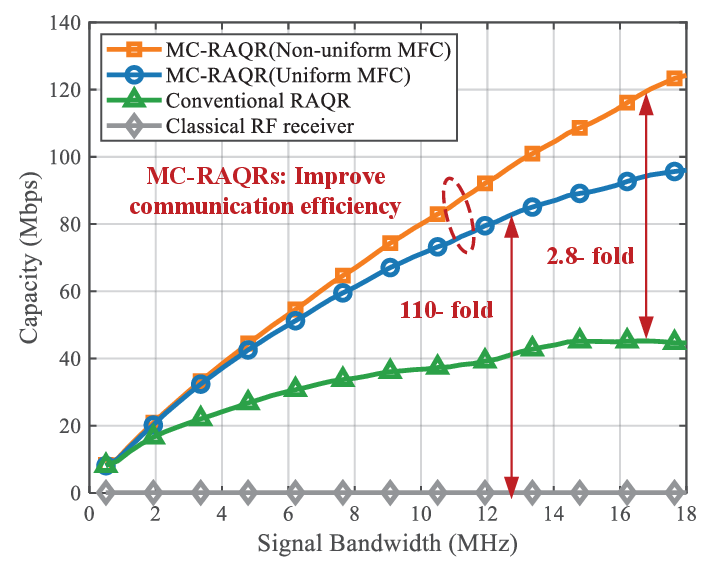}}
		\vspace{0.4em}
	\caption{\textcolor{black}{Comparison of the communication capacity versus signal bandwidth for the proposed MC-RAQR schemes, conventional RAQRs, and the classical RF receivers. The distance between transmitter and receiver is $1500$m. }}
	\label{com}
	\vspace{-1.5ex}
\end{figure}
\textcolor{black}{In this section, we compare the communication and sensing performance of the proposed MC-RAQRs to both the conventional RAQRs and classical RF receivers. More particularly, classical RF receivers are typically composed of an antenna, an LNA, and the remaining components such as mixers, lowpass filters, and ADCs, where their corresponding gains are $G_{\text{ANT}}$, $G_{\text{LNA}}$, and $G_{\text{REC}}$, respectively. The equivalent baseband signal model of the classical RF transceiver system can be formulated as a similar form as (17), where $\kappa_m$ becomes  ${\kappa _m} = \sqrt {{A_e}{\eta _0}{G_{\text{ANT}}}{G_{\text{LNA}}}{G_{\text{REC}}}}$, ${A_e} = {\lambda ^2}/4\pi$ denotes the effective aperture of an isotropic antenna, and $\eta_0$ is the antenna efficiency. The noise of classical RF receivers is assumed to be AWGN with variance $\sigma _{RF}^2 = {k_B}TB{G_{\text{LNA}}} F$, where $ F$ represents the noise factor of the holistic receiver system. Without loss of generality, we assume $\eta_0=0.7$, ${G_{\text{ANT}}} = 5.5$ dB, ${G_{\text{REC}}} = 1$, and ${ F}=6$ dB \cite{3gpp}.}

Fig. \ref{com} portrays the capacity of four considered schemes when transmitting the multi-carrier communication signal. The transmission distance is set to be $r=1500$ m. It is observed that utilizing our MC-RAQR schemes can achieve up to 110-fold higher capacity, compared to the employment of classical RF receivers. Furthermore, the ``Non-Uniform MFC" scheme obtains the best performance due to well-designed MFC structure and larger $\kappa$, while the capacity of the ``Uniform-MFC" scheme is slightly lower. The capacity of conventional RAQRs with a single LO is close to the former two schemes when the signal bandwidth is small, but shows diminishing return as the bandwidth increases. This is reasonable because of smaller 3dB bandwidth of the ``conventional RAQR" scheme, as portrayed in Fig. \ref{3dbw}.
\begin{figure}[!tbp]
	\setlength{\abovecaptionskip}{-0.1cm}
	\setlength{\belowcaptionskip}{-0.1cm}
	\centerline{\includegraphics[width=0.45\textwidth]{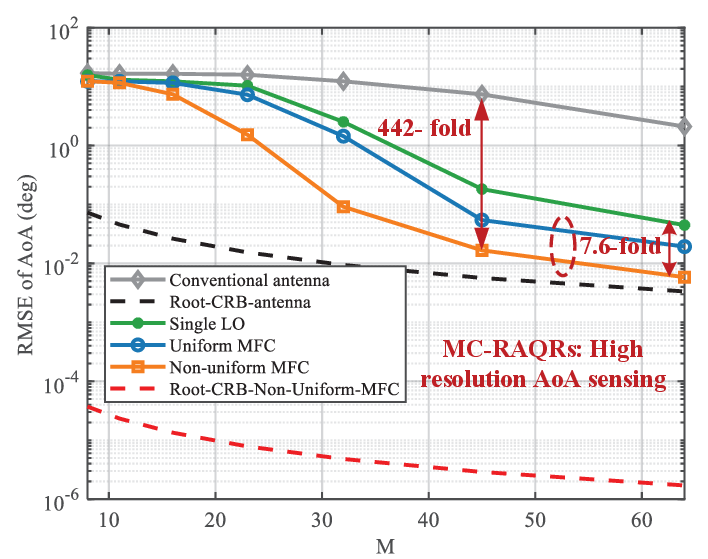}}
		\vspace{0.4em}
	\caption{\textcolor{black}{RMSE of the AoA estimation versus the number of sensors. The proposed uniform and non-uniform MC-RAQR schemes significantly outperform the conventional RAQRs and classical RF receivers.}}
	\label{sen_M}
	\vspace{-1.5ex}
\end{figure}

\textcolor{black}{Fig. \ref{sen_M} shows the accuracy of AoA estimation as the number of sensor increases, where $N=10$ subcarriers are considered and four close targets located at ($16.1^\circ$, $300.1$m), ($19.4^\circ$, $330.2$m), ($23.5^\circ$, $370.3$m), and ($26.9^\circ$, $400.4$m) are evaluated. The proposed MC-RAQR schemes, conventional RAQRs with single LO, and the classical RF receivers are plotted for comparison. Additionally, the root-CRBs for classical RF receivers and the proposed MC-RAQR with non-uniform MFC are also presented. It is observed that the accuracy of the AoA estimation increases with number of sensors $M$, thanks to the enlarged effective aperture. The classical RF receivers have the worst performance due to low receive SNR, while conventional RAQRs and proposed MC-RAQR schemes benefit from high receiver gain. Furthermore, the CRB of the MC-RAQR is much lower than that of the classical RF receivers. More particularly, the proposed MC-RAQR with non-uniform MFC achieves the lowest RMSE for large $M$, yielding a  $442$-fold and $7.6$-fold reduction in the RMSE compared to the classical RF receivers and conventional RAQR schemes, respectively.}

\begin{figure}[tbp]
	\setlength{\abovecaptionskip}{-0.1cm}
	\setlength{\belowcaptionskip}{-0.1cm}
	\centerline{\includegraphics[width=0.45\textwidth]{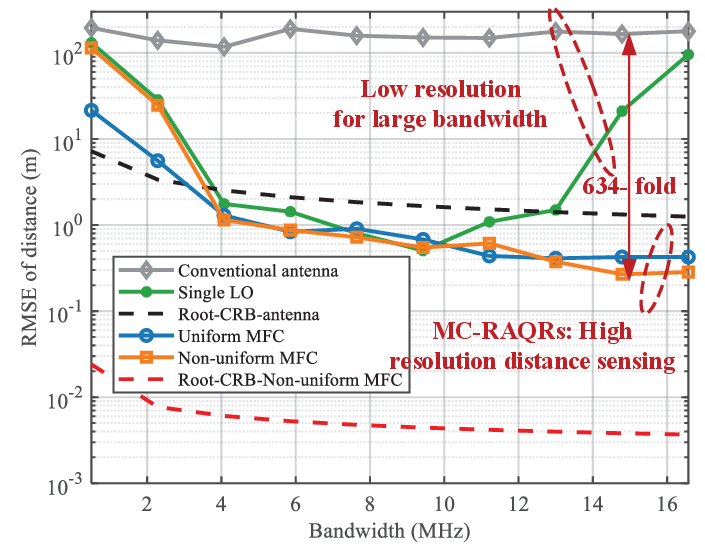}}
		\vspace{0.4em}	
	\caption{\textcolor{black}{RMSE of distance estimation versus the RF signal bandwidth.}}
	\label{sen_N}
	\vspace{-1.5ex}
\end{figure}
\textcolor{black}{Fig. \ref{sen_N} shows the RMSE of the distance estimate versus the bandwidth. The number of sensors is set to be $M=40$ and the locations of four targets are ($16.1^\circ$, $580.1$m), ($19.4^\circ$, $630.2$m), ($23.5^\circ$, $660.3$m), and ($26.9^\circ$, $710.4$m). It is observed that the accuracy of distance estimation increases gradually with the bandwidth for MC-RAQR schemes, thanks to the increasing bandwidth. However, the RMSE remains large for classical RF receivers even exploiting large bandwidth. This performance degradation is primarily due to error propagation. Since the distance calculation fundamentally relies on the AoA, the inaccuracies in the AoA estimation severely compromise the subsequent distance estimation. For conventional RAQRs with a single LO, their RMSE is similar to the proposed MC-RAQR schemes for relatively small bandwidth, but drastically deteriorate when the bandwidth exceeds $10$ MHz, demonstrating the performance degradation of conventional RAQRs in wideband signal reception. Comparing the CRB between classical RF receivers and the proposed MC-RAQR with non-uniform MFC, we observe that our MC-RAQR has a much lower CRB. More particularly, the proposed MC-RAQR schemes can achieve the RMSE $634$-fold smaller than that of the classical RF receivers, indicating their remarkable accuracy and performance superiority. This result is consistent with the theoretical analysis of distance resolution. Specifically, based on the definition of distance resolution \cite{crb1}, the conventional bandwidth-limited RAQRs (using a single LO) yields a coarse distance resolution of $214$ m under the condition of their $\sim 700$ kHz bandwidth (as seen in Fig. 6). This indicates that targets separated by less than $214$ meters cannot be effectively distinguished by employing the conventional RAQRs, which accounts for their severe RMSE degradation shown in Fig. \ref{sen_N}. Conversely, the distance resolution is dramatically refined to about $\sim 25$ m for the proposed MC-RAQR scheme thanks to an extended bandwidth to $\sim 12$ MHz.}

\subsection{\textcolor{black}{Discussion on impairment}}
\begin{figure*}[h!]
	\hspace{-1.1cm}
	\subfloat[ Impact of CFO.]{
		\includegraphics[width=0.37\textwidth]{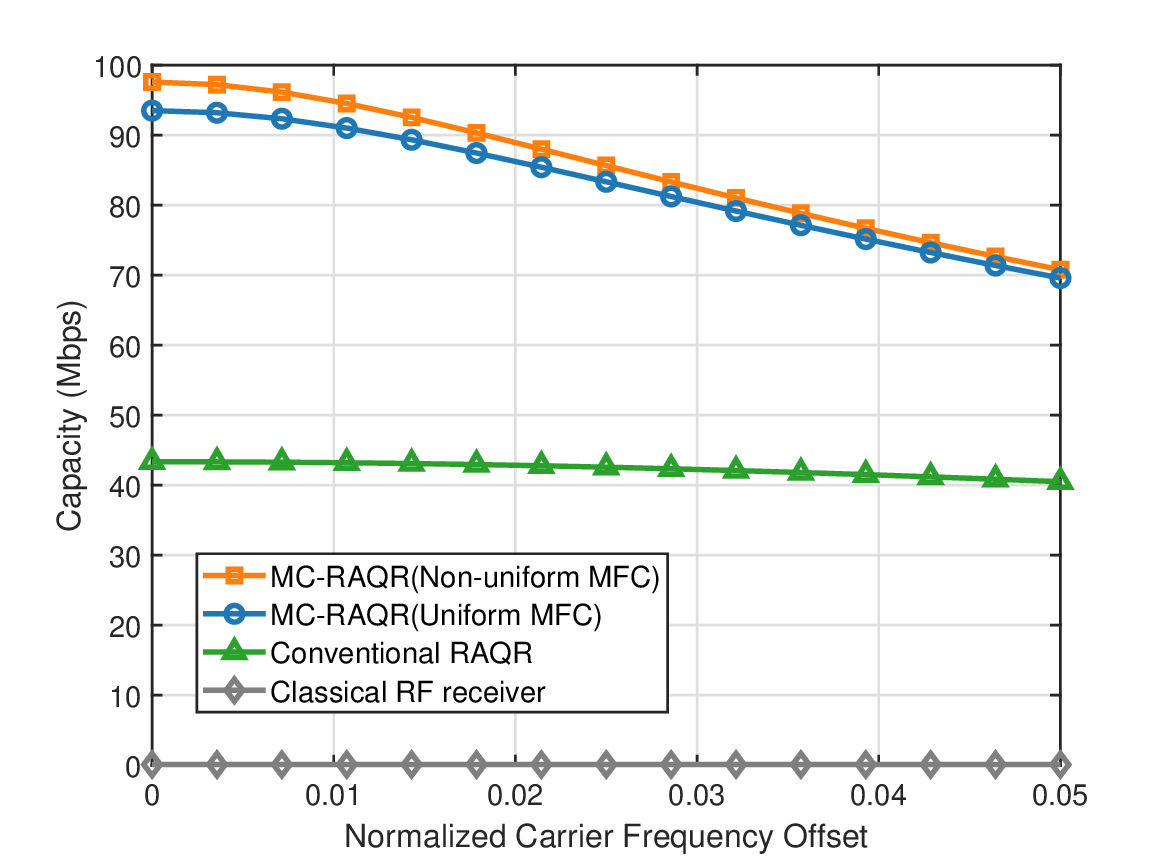}}
	\hspace{-0.5cm}
	\subfloat[Impact of RF Phase Noise.]{
		\includegraphics[width=0.37\textwidth]{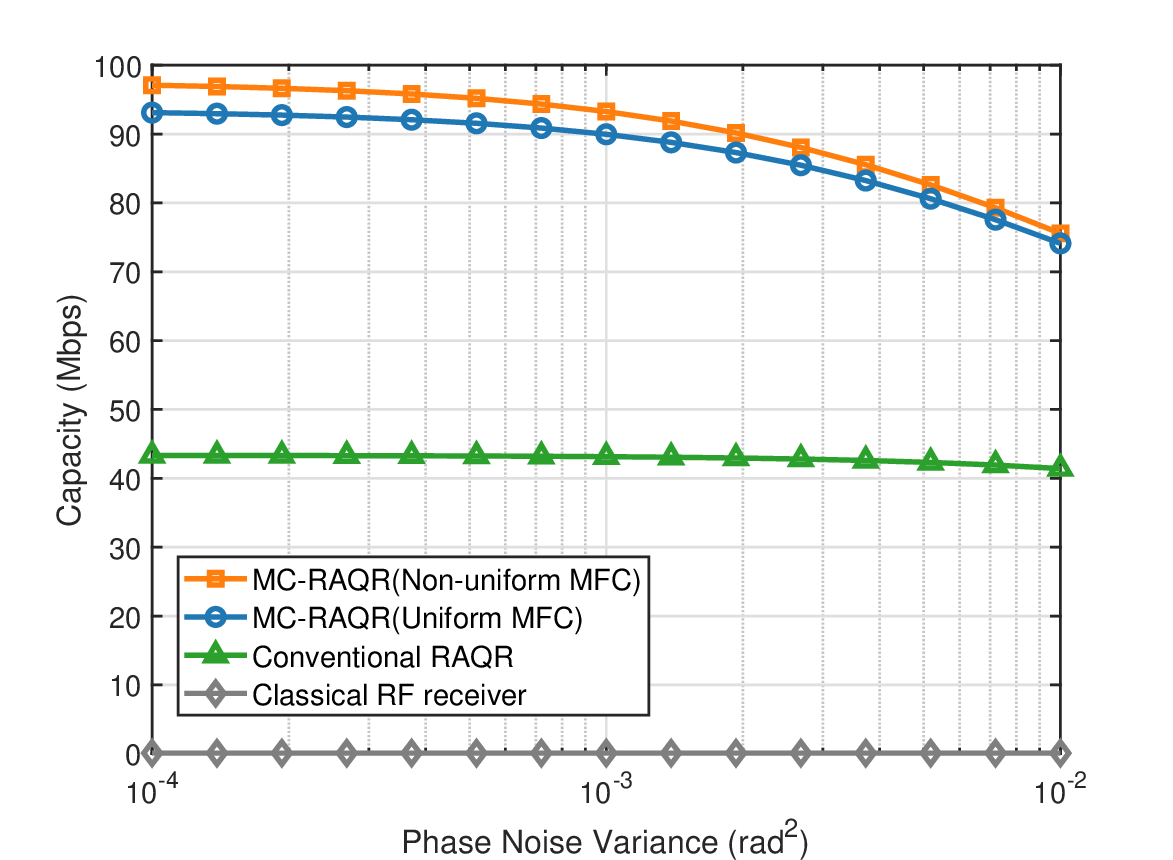}} 
	\hspace{-0.5cm}
	\subfloat[Impact of Optical Phase Shift.]{
		\includegraphics[width=0.37\textwidth]{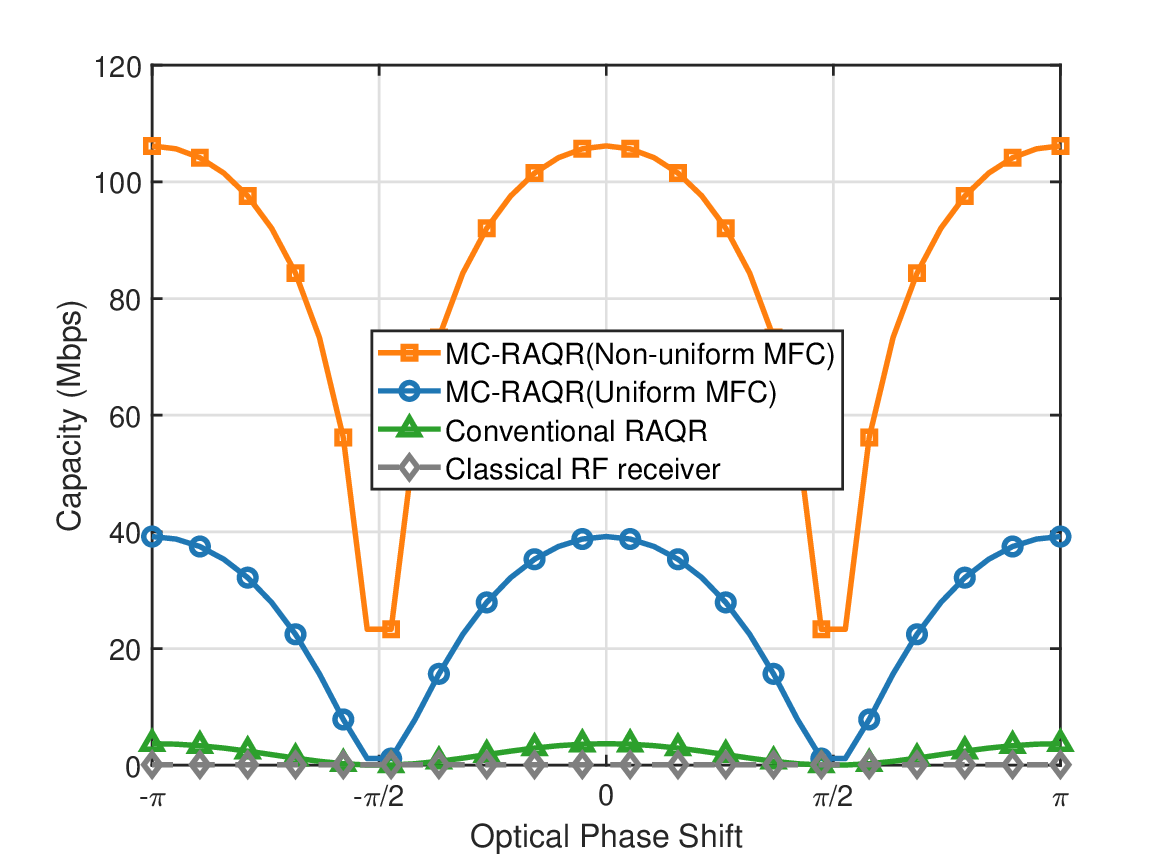}}
	\vspace{-0.5em}
	\caption{\textcolor{black}{Robustness evaluation of the communication capacity under practical system impairments. The proposed MC-RAQR architectures demonstrate consistent superiority over the conventional baselines under varying practical conditions.}}
	\label{impairments}
\end{figure*}

\textcolor{black}{In this section, we evaluate the performance of proposed MC-RAQR under several practical impairments, such as carrier frequency offset (CFO), phase noise, and optical phase shift. Without loss of generality, the bandwidth in Fig. \ref{impairments} is set to be $12$ MHz. Fig. \ref{impairments} (a) shows the influence of the CFO between the transmitter and receiver, namely the carrier frequency mismatch between the LO and the transmit signal \cite{LO}. It is observed that the capacity of the proposed MC-RAQRs decreases gradually with the normalized CFO in contrast to the conventional RAQRs and classical RF receivers. This is reasonable because the MFC architecture employed by our MC-RAQR successfully aligns the optical phases across the multi-carrier signal to achieve a highly coherent detection, hence, it inherently becomes more sensitive to phase perturbations. It is also worth noting that the proposed MC-RAQR scheme maintains a significant higher capacity compared to both the convention RAQRs and classical RF receivers, even under severe CFOs. In practical deployments, this residual impairment can be further mitigated using standard digital synchronization algorithms.}

\textcolor{black}{Fig. \ref{impairments} (b) demonstrates the capacity sensitivity to RF phase noise, where the performance loss stems from the combined effects of common phase error (CPE) and additional ICI \cite{phase_noise}. It is obvious that the proposed MC-RAQR architectures exhibit high robustness against relatively small RF phase noise, maintaining a stable capacity profile. Although a gradual degradation of the capacity is observed as the phase noise variance becomes larger, the proposed scheme still maintains a substantial performance advantage. More particularly, even under severe phase noise conditions with a variance of $10^{-2} \text{ rad}^2$, the capacity of our MC-RAQR remains approximately 2 times and 80 times higher than that of the conventional RAQRs and of the classical RF receivers, respectively.}

\textcolor{black}{Finally, we evaluate the impact of the optical phase mismatch in the BCOD scheme. The output of the BCOD is inherently governed by a cosine interference envelope as characterized in (\ref{output_PD}), hence the proposed MC-RAQR scheme exhibits obvious interference patterns with deep nulls at $\pm\pi/2$ and maximum capacities at $0$ and $\pm\pi$, as shown in Fig. \ref{impairments} (c). By contrast, the conventional RAQR scheme shows a relatively flattened response with negligible interference, indicating that the conventional RAQR scheme suffers from severe atomic dispersive dephasing under multi-carrier RF signals, which further destroys the coherence of carrier phases. By effectively mapping all subcarriers to low IFs, the proposed MC-RAQR scheme successfully restores phase coherence for multi-carrier signals, hence enabling a highly efficient coherent detection.}

\section{Conclusion}\label{conclusion}
This paper proposed a five-level MC-RAQR structure for realizing the multi-carrier communication and sensing. We first developed a five-level quantum system model and derived the amplitude and phase of probe laser as well as the output electrical signal for signal processing. Performance evaluations were then conducted for both multi-carrier communication and multi-target sensing, including the derivation of channel capacity and CRBs of angle and range estimation. Numerical simulations validated the proposed scheme, model and algorithms, demonstrating that the proposed MC-RAQR scheme achieved approximately a bandwidth of $11.7$ MHz, significantly enhanced communication channel capacity, and enabled high-resolution multi-target sensing, compared to conventional RAQRs and antennas. Thus highlighting its strong potential and advantages in multi-carrier signal reception.

\section*{Appendix A: Proof of Theorem \ref{Uzm}}
Based on (\ref{xm}) and (\ref{ym}), the superimposed RF signal $z_m(t)$ can be expressed as
\begin{equation}
	\setlength\abovedisplayskip{2.7pt}
	\setlength\belowdisplayskip{2.7pt}
	\small
	\begin{aligned}
	{z_m}(t) = \sqrt 2 {\cal R} \left\{ {\sum\limits_{i = 0}^{N-1} {{x_{i,m}}(t){e^{j2\pi {f_{x,i}}t}}}  + \sum\limits_{j = 1}^B {{y_{j,m}}(t){e^{j2\pi {f_{y,j,m}}t}}} } \right\}.
	\end{aligned}
\end{equation}
The power of $z_m(t)$ is derived as 
\begin{equation}
	\setlength\abovedisplayskip{2.7pt}
	\setlength\belowdisplayskip{2.7pt}
\begin{aligned}
	{P_{z,m}} &= {\left( {\sum\limits_{i = 0}^{N-1} {{x_{i,m}}(t){e^{j2\pi {f_{x,i}}t}}}  + \sum\limits_{j = 1}^B {{y_{j,m}}(t){e^{j2\pi {f_{y,j,m}}t}}} } \right)^2}\\
	&\mathop  \approx \limits^{(a)} \sum\limits_{i = 0}^{N-1} {{{\mathcal{P}}_{x,i,m}}}  + \sum\limits_{j = 1}^B {\frac{{\mathcal{P}}_{y,m}}{B}}  + \sum\limits_{i = 0}^{N-1} {\sum\limits_{j = 1}^B {\sqrt {{{\mathcal{P}}_{x,i,m}}} } } \sqrt {\frac{{\mathcal{P}}_{y,m}}{B}} \\
	&\times {e^{j\Delta {\phi _{i,j,m}}}}{e^{j2\pi \Delta {f_{i,j,m}}t}} + \sum\limits_{i = 0}^{N-1} {\sum\limits_{j = 1}^B {\sqrt {{{\mathcal{P}}_{x,i,m}}} } } \sqrt {\frac{{\mathcal{P}}_{y,m}}{B}} \\
	&\times {e^{ - j\Delta {\phi _{i,j,m}}}}{e^{ - j2\pi\Delta {f_{i,j,m}}t}}\\
	&= \sum\limits_{i = 0}^{N-1} {{{\mathcal{P}}_{x,i,m}}}  + {{\mathcal{P}}_{y,m}} + 2\sum\limits_{i = 0}^{N-1} {\sum\limits_{j = 1}^B {\sqrt {{{\mathcal{P}}_{x,i,m}}} } } \sqrt {\frac{{\mathcal{P}}_{y,m}}{B}} \\
	&\times\cos (\Delta {\phi _{i,j,m}} + 2\pi \Delta {f_{i,j,m}}t),
\end{aligned}
\end{equation}
where $\Delta {f_{i,j,m}} = {f_{x,i}} - {f_{y,j,m}}$ and $\Delta {\phi _{i,j,m}} = {\theta _{i,m}} - {\varphi _{j,m}}$; $(a)$ is obtained by assuming that the frequency interference terms between the multi-carrier signal and MFC lines can be neglected. To derive the expression of $U_{z,m}$, we first calculate $\sqrt{P_{z,m}}$ as follows 
\begin{equation}
	\setlength\abovedisplayskip{2.7pt}
	\setlength\belowdisplayskip{2.7pt}
\begin{aligned}
	&\sqrt {{P_{z,m}}}  = \sqrt {{{\mathcal{P}}_{y,m}}} \sqrt {\begin{array}{l}
		1 + \frac{{\sum\limits_{i = 0}^{N-1} {{{\mathcal{P}}_{x,i,m}}} }}{{{{\mathcal{P}}_{y,m}}}} + 2\sum\limits_{i = 0}^{N-1} {\sum\limits_{j = 1}^B {\sqrt {\frac{{{{\mathcal{P}}_{x,i,m}}}}{{B{{\mathcal{P}}_{y,m}}}}} } } \\
		\cos (2\pi \Delta {f_{i,j,m}}t + \Delta {\phi _{i,j,m}})
	\end{array}} \\
	&\mathop  \approx \limits^{(b)} \sqrt {{{\mathcal{P}}_{y,m}}}  + \frac{1}{{\sqrt B }}\sum\limits_{i = 0}^{N-1} {\sqrt {{{\mathcal{P}}_{x,i,m}}} \sum\limits_{j = 1}^B {\cos (2\pi \Delta {f_{i,j,m}}t + \Delta {\phi _{i,j,m}})} } \\
	&\mathop  \approx \limits^{(c)} \sqrt {{{\mathcal{P}}_{y,m}}}  + \frac{1}{{\sqrt B }}\sum\limits_{i = 0}^{N-1} {\sqrt {{{\mathcal{P}}_{x,i,m}}} \cos (2\pi \Delta {f_{ni,m}}t + \Delta {\phi _{ni,m}})},
\end{aligned}
\end{equation}
where $(b)$ is obtained by assuming the multi-carrier signals are weak compared with the MFC signal, i.e., $\sum_{i = 0}^{N-1} {{{\mathcal{P}}_{x,i,m}}}  \ll {{\mathcal{P}}_{y,m}}.$ $(c)$ is obtained because for each carrier frequency $i$, by setting the MFC repetition rate larger than the instantaneous bandwidth, the strongest beat signal comes from its nearest MFC comb line, and other beat signals are negligible \cite{comb}. Thus, we have $\sum_{j = 1}^B {\cos (2\pi \Delta {f_{i,j,m}}t + \Delta {\phi _{i,j,m}})}  \approx \cos (2\pi \Delta {f_{ni,m}}t + \Delta {\phi _{ni,m}}), \forall i$, where $ni$ is the index of the nearest comb line for frequency $i$. Thus, the amplitude of the superposition signal $z_m(t)$ is 
\begin{equation}
	\setlength\abovedisplayskip{2.7pt}
	\setlength\belowdisplayskip{2.7pt}
\begin{aligned}
	{U_{z,m}} &= \sqrt {\frac{2}{{c{_0}{A_e}}}} \sqrt {{P_{z,m}}} \\
	&=  {U_{y,m}} + \frac{1}{{\sqrt B }}\sum\limits_{i = 0}^{N-1} {{U_{x,i,m}}\cos (2\pi \Delta {f_{ni,m}}t + \Delta {\phi _{ni,m}})}. 
\end{aligned}
\end{equation}

\section*{Appendix B: The coefficients in (\ref{rho21})}
The coefficients of $\rho_{21}$ are derived by solving the steady state of the Lindblad master equation, and are given by 
\allowdisplaybreaks
\begin{align}
	\overline{o}_1 &=  - 2{\Delta _c}\Omega _c^2{\Omega _p}. \\
	\overline{o}_2 &= 2\Omega _c^2{\Omega _p}( - 8\,{\Delta _c^2} + 8\,{\Delta _a}\,{\Delta _c} + {\Omega _a^2})({\Delta _a} - {\Delta _c} + {\Delta _x}).\\
	\overline{o}_3 &=- 8\Omega _c^2{\Omega _p}({\Delta _a} - {\Delta _c})( - 4\Delta _c^2 + 4{\Delta _a}{\Delta _c} + \Omega _a^2)\nonumber\\
	&\times{({\Delta _a} - {\Delta _c} + {\Delta _x})^2}.\nonumber\\ 
	\overline{o}_4 &= 4\Delta _c^2{\Omega _p}{\gamma _2}.\\
	\overline{o}_5& =  - 8{\Delta _c}{\Omega _p}{\gamma _2}( - 4\Delta _c^2 + 4{\Delta _a}{\Delta _c} + \Omega _a^2)({\Delta _a} - {\Delta _c} + {\Delta _x}). \\
	\overline{o}_6& = 4{\Omega _p}{\gamma _2}{( - 4\Delta _c^2 + 4{\Delta _a}{\Delta _c} + \Omega _a^2)^2}{({\Delta _a} - {\Delta _c} + {\Delta _x})^2}.\\
	\underline{o}_1& = 8{\mkern 1mu} {\Delta _c^2}{\mkern 1mu} {\Omega _p^2} + 4{\mkern 1mu} {\Delta _c^2}{\mkern 1mu} {\gamma _2^2} + {\Omega _c^4} + 2{\mkern 1mu} {\Omega _c^2}{\mkern 1mu} {\Omega _p^2} + {\Omega _p^4}.\\
	\underline{o}_2& = - 2(32{\mkern 1mu} {\Delta _a^2}{\mkern 1mu} {\Delta _c^2}{\mkern 1mu} {\Omega _p^2} + 16{\mkern 1mu} {\Delta _a^2}{\mkern 1mu} {\Delta _c^2}{\mkern 1mu} {\gamma _2^2} + 4{\mkern 1mu} {\Delta _a^2}{\mkern 1mu} {\Omega _c^4}\nonumber\\
	& + 8{\mkern 1mu} {\Delta _a^2}{\mkern 1mu} {\Omega _c^2}{\mkern 1mu} {\Omega _p^2} + 4{\mkern 1mu} {\Delta _a^2}{\mkern 1mu} {\Omega _p^4} - 64{\mkern 1mu} {\Delta _a}{\mkern 1mu} {\Delta _c^3}{\mkern 1mu} {\Omega _p^2} - {\Omega _a^2}{\mkern 1mu} {\Omega _p^4} \nonumber\\
	&- 32{\mkern 1mu} {\Delta _a}{\mkern 1mu} {\Delta _c^3}{\mkern 1mu} {\gamma _2^2} + 32{\mkern 1mu} {\Delta _x}{\mkern 1mu} {\Delta _a}{\mkern 1mu} {\Delta _c^2}{\mkern 1mu} {\Omega _p^2} + 16{\mkern 1mu} {\Delta _x}{\mkern 1mu} {\Delta _a}{\mkern 1mu} {\Delta _c^2}{\mkern 1mu} {\gamma _2^2} \nonumber\\
	&+ 8{\mkern 1mu} {\Delta _a}{\mkern 1mu} {\Delta _c}{\mkern 1mu} {\Omega _a^2}{\mkern 1mu} {\Omega _p^2} + 4{\mkern 1mu} {\Delta _a}{\mkern 1mu} {\Delta _c}{\mkern 1mu} {\Omega _a^2}{\mkern 1mu} {\gamma _2^2} - 8{\mkern 1mu} {\Delta _a}{\mkern 1mu} {\Delta _c}{\mkern 1mu} {\Omega _c^4} \nonumber\\
	&- 16{\mkern 1mu} {\Delta _a}{\mkern 1mu} {\Delta _c}{\mkern 1mu} {\Omega _c^2}{\mkern 1mu} {\Omega _p^2} - 8{\mkern 1mu} {\Delta _a}{\mkern 1mu} {\Delta _c}{\mkern 1mu} {\Omega _p^4} + 4{\mkern 1mu} {\Delta _x}{\mkern 1mu} {\Delta _a}{\mkern 1mu} {\Omega _c^4} \nonumber\\
	&+ 8{\mkern 1mu} {\Delta _x}{\mkern 1mu} {\Delta _a}{\mkern 1mu} {\Omega _c^2}{\mkern 1mu} {\Omega _p^2} + 4{\mkern 1mu} {\Delta _x}{\mkern 1mu} {\Delta _a}{\mkern 1mu} {\Omega _p^4} + 32{\mkern 1mu} {\Delta _c^4}{\mkern 1mu} {\Omega _p^2} \nonumber\\
	&+ 16{\mkern 1mu} {\Delta _c^4}{\mkern 1mu} {\gamma _2^2} - 32{\mkern 1mu} {\Delta _x}{\mkern 1mu} {\Delta _c^3}{\mkern 1mu} {\Omega _p^2} - 16{\mkern 1mu} {\Delta _x}{\mkern 1mu} {\Delta _c^3}{\mkern 1mu} {\gamma _2^2} \nonumber\\
	&- 8{\mkern 1mu} {\Delta _c^2}{\mkern 1mu} {\Omega _a^2}{\mkern 1mu} {\Omega _p^2} - 4{\mkern 1mu} {\Delta _c^2}{\mkern 1mu} {\Omega _a^2}{\mkern 1mu} {\gamma _2^2} + 4{\mkern 1mu} {\Delta _c^2}{\mkern 1mu} {\Omega _c^4}- {\Omega _a^2}{\mkern 1mu} {\Omega _c^2}{\mkern 1mu} {\Omega _p^2} \nonumber\\
	&+ 8{\mkern 1mu} {\Delta _c^2}{\mkern 1mu} {\Omega _c^2}{\mkern 1mu} {\Omega _p^2} + 8{\mkern 1mu} {\Delta _x}{\mkern 1mu} {\Delta _c}{\mkern 1mu} {\Omega _a^2}{\mkern 1mu} {\Omega _p^2} + 4{\mkern 1mu} {\Delta _x}{\mkern 1mu} {\Delta _c}{\mkern 1mu} {\Omega _a^2}{\mkern 1mu} {\gamma _2^2} \nonumber\\
	&- 4{\mkern 1mu} {\Delta _x}{\mkern 1mu} {\Delta _c}{\mkern 1mu} {\Omega _c^4} - 8{\mkern 1mu} {\Delta _x}{\mkern 1mu} {\Delta _c}{\mkern 1mu} {\Omega _c^2}{\mkern 1mu} {\Omega _p^2} - 4{\mkern 1mu} {\Delta _x}{\mkern 1mu} {\Delta _c}{\mkern 1mu} {\Omega _p^4}.\\
	\underline{o}_3& = 128\,{\Delta_a ^4} \,{\Delta_c ^2}  {\Omega_p ^2} +64 {\Delta_a ^4}  {\Delta_c ^2}  {\gamma_2 ^2} +16 {\Delta_a ^4}  {\Omega_c ^4} \nonumber\\
	&+32 {\Delta_a ^4}  {\Omega_c ^2}  {\Omega_p ^2} +16 {\Delta_a ^4}  {\Omega_p ^4} -512 {\Delta_a ^3}  {\Delta_c ^3}  {\Omega_p ^2}  \nonumber\\
	&+256 {\Delta_a ^3}  {\Delta_c ^2}  \Delta_s  {\Omega_p ^2} +128 {\Delta_a ^3}  {\Delta_c ^2}  \Delta_s  {\gamma_2 ^2} \nonumber\\
	&+64 {\Delta_a ^3}  \Delta_c  {\Omega_a ^2}  {\Omega_p ^2} -256 {\Delta_a ^3}  {\Delta_c ^3}  {\gamma_2 ^2}+768 {\Delta_a ^2}  {\Delta_c ^4}  {\Omega_p ^2} \nonumber\\
	&-64 {\Delta_a ^3}  \Delta_c  {\Omega_c ^4} -128 {\Delta_a ^3}  \Delta_c  {\Omega_c ^2}  {\Omega_p ^2}+384 {\Delta_a ^2}  {\Delta_c ^4}  {\gamma_2 ^2}  \nonumber\\
	&+32 {\Delta_a ^3}  \Delta_s  {\Omega_c ^4} +64 {\Delta_a ^3}  \Delta_s  {\Omega_c ^2}  {\Omega_p ^2} +32 {\Delta_a ^3}  \Delta_s  {\Omega_p ^4} \nonumber\\
	&-384 {\Delta_a ^2}  {\Delta_c ^3}  \Delta_s  {\gamma_2 ^2} +128 {\Delta_a ^2}  {\Delta_c ^2}  {\Delta_s ^2}  {\Omega_p ^2} \nonumber\\
	&+64 {\Delta_a ^2}  {\Delta_c ^2}  {\Delta_s ^2}  {\gamma_2 ^2} -192 {\Delta_a ^2}  {\Delta_c ^2}  {\Omega_a ^2}  {\Omega_p ^2} \nonumber\\
	&-96 {\Delta_a ^2}  {\Delta_c ^2}  {\Omega_a ^2}  {\gamma_2 ^2} +96 {\Delta_a ^2}  {\Delta_c ^2}  {\Omega_c ^4} +{\Omega_a ^4}  {\Omega_p ^4}\nonumber\\
	&+192 {\Delta_a ^2}  {\Delta_c ^2}  {\Omega_c ^2}  {\Omega_p ^2} +128 {\Delta_a ^2}  {\Delta_c ^2}  {\Omega_p ^4} \nonumber\\
	&+128 {\Delta_a ^2}  \Delta_c  \Delta_s  {\Omega_a ^2}  {\Omega_p ^2} +64 {\Delta_a ^2}  \Delta_c  \Delta_s  {\Omega_a ^2}  {\gamma_2 ^2} \nonumber\\
	&-96 {\Delta_a ^2}  \Delta_c  \Delta_s  {\Omega_c ^4} -192 {\Delta_a ^2}  \Delta_c  \Delta_s  {\Omega_c ^2}  {\Omega_p ^2} \nonumber\\
	&-96 {\Delta_a ^2}  \Delta_c  \Delta_s  {\Omega_p ^4} +16 {\Delta_a ^2}  {\Delta_s ^2}  {\Omega_c ^4} +32 {\Delta_a ^2}  {\Delta_s ^2}  {\Omega_c ^2}  {\Omega_p ^2} \nonumber\\
	&+16 {\Delta_a ^2}  {\Delta_s ^2}  {\Omega_p ^4} +8 {\Delta_a ^2}  {\Omega_a ^4}  {\Omega_p ^2} +4 {\Delta_a ^2}  {\Omega_a ^4}  {\gamma_2 ^2}\nonumber\\
	& +8 {\Delta_a ^2}  {\Omega_a ^2}  {\Omega_c ^2}  {\Omega_p ^2} +8 {\Delta_a ^2}  {\Omega_a ^2}  {\Omega_p ^4} -512 \Delta_a  {\Delta_c ^4}  {\Omega_p ^2}\nonumber\\
	& -256 \Delta_a  {\Delta_c ^4}  {\gamma_2 ^2} +768 \Delta_a  {\Delta_c ^4}  \Delta_s  {\Omega_p ^2}  \nonumber\\
	&-256 \Delta_a  {\Delta_c ^3}  {\Delta_s ^2}  {\Omega_p ^2} -128 \Delta_a  {\Delta_c ^3}  {\Delta_s ^2}  {\gamma_2 ^2} \nonumber\\
	&+192 \Delta_a  {\Delta_c ^3}  {\Omega_a ^2}  {\Omega_p ^2} +96 \Delta_a  {\Delta_c ^3}  {\Omega_a ^2}  {\gamma_2 ^2}  \nonumber\\
	&-128 \Delta_a  {\Delta_c ^3}  {\Omega_c ^2}  {\Omega_p ^2} -128 \Delta_a  {\Delta_c ^3}  {\Omega_p ^4} \nonumber\\
	&-256 \Delta_a  {\Delta_c ^2}  \Delta_s  {\Omega_a ^2}  {\Omega_p ^2} -128 \Delta_a  {\Delta_c ^2}  \Delta_s  {\Omega_a ^2}  {\gamma_2 ^2} \nonumber\\
	&+96 \Delta_a  {\Delta_c ^2}  \Delta_s  {\Omega_c ^4} +192 \Delta_a  {\Delta_c ^2}  \Delta_s  {\Omega_c ^2}  {\Omega_p ^2} \nonumber\\
	&+128 \Delta_a  {\Delta_c ^2}  \Delta_s  {\Omega_p ^4} +64 \Delta_a  \Delta_c  {\Delta_s ^2}  {\Omega_a ^2}  {\Omega_p ^2} \nonumber\\
	&+32 \Delta_a  \Delta_c  {\Delta_s ^2}  {\Omega_a ^2}  {\gamma_2 ^2} -32 \Delta_a  \Delta_c  {\Delta_s ^2}  {\Omega_c ^4} \nonumber\\
	&-64 \Delta_a  \Delta_c  {\Delta_s ^2}  {\Omega_c ^2}  {\Omega_p ^2} -32 \Delta_a  \Delta_c  {\Delta_s ^2}  {\Omega_p ^4} \nonumber\\
	&-16 \Delta_a  \Delta_c  {\Omega_a ^4}  {\Omega_p ^2} -8 \Delta_a  \Delta_c  {\Omega_a ^4}  {\gamma_2 ^2}+384 \Delta_a  {\Delta_c ^4}  \Delta_s  {\gamma_2 ^2}  \nonumber\\
	&-8 \Delta_a  \Delta_c  {\Omega_a ^2}  {\Omega_p ^4} +16 \Delta_a  \Delta_s  {\Omega_a ^4}  {\Omega_p ^2} +8 \Delta_a  \Delta_s  {\Omega_a ^4}  {\gamma_2 ^2} \nonumber\\
	&+16 \Delta_a  \Delta_s  {\Omega_a ^2}  {\Omega_c ^2}  {\Omega_p ^2} +16 \Delta_a  \Delta_s  {\Omega_a ^2}  {\Omega_p ^4}  \nonumber\\
	&+64 {\Delta_c }^6  {\gamma_2 ^2} -256 {\Delta_c ^4}  \Delta_s  {\Omega_p ^2} -128 {\Delta_c ^4}  \Delta_s  {\gamma_2 ^2} \nonumber\\
	&+128 {\Delta_c ^4}  {\Delta_s ^2}  {\Omega_p ^2} +64 {\Delta_c ^4}  {\Delta_s ^2}  {\gamma_2 ^2} -64 {\Delta_c ^4}  {\Omega_a ^2}  {\Omega_p ^2} \nonumber\\
	&-32 {\Delta_c ^4}  {\Omega_a ^2}  {\gamma_2 ^2} +16 {\Delta_c ^4}  {\Omega_c ^4}+16 {\Delta_c ^2}  {\Delta_s ^2}  {\Omega_c ^4}   \nonumber\\
	&+128 {\Delta_c ^3}  \Delta_s  {\Omega_a ^2}  {\Omega_p ^2} +64 {\Delta_c ^3}  \Delta_s  {\Omega_a ^2}  {\gamma_2 ^2}  \nonumber\\
	&-64 {\Delta_c ^3}  \Delta_s  {\Omega_c ^2}  {\Omega_p ^2} -64 {\Delta_c ^3}  \Delta_s  {\Omega_p ^4}-32 {\Delta_c ^2}  {\Delta_s ^2}  {\Omega_a ^2}  {\gamma_2 ^2}  \nonumber\\
	&+32 {\Delta_c ^2}  {\Delta_s ^2}  {\Omega_p ^4} +8 {\Delta_c ^2}  {\Omega_a ^4}  {\Omega_p ^2} +4 {\Delta_c ^2}  {\Omega_a ^4}  {\gamma_2 ^2} \nonumber\\
	&+8 {\Delta_c ^2}  {\Omega_a ^2}  {\Omega_c ^2}  {\Omega_p ^2} -16 \Delta_c  \Delta_s  {\Omega_a ^4}  {\Omega_p ^2}-64 \Delta_a  {\Delta_c ^3}  {\Omega_c ^4}  \nonumber\\
	&-16 \Delta_c  \Delta_s  {\Omega_a ^2}  {\Omega_c ^2}  {\Omega_p ^2} -16 \Delta_c  \Delta_s  {\Omega_a ^2}  {\Omega_p ^4}  \nonumber\\
	&+4 {\Delta_s ^2}  {\Omega_a ^4}  {\gamma_2 ^2} +8 {\Delta_s ^2}  {\Omega_a ^2}  {\Omega_c ^2}  {\Omega_p ^2} +8 {\Delta_s ^2}  {\Omega_a ^2}  {\Omega_p ^4}\nonumber\\
	&+32 {\Delta_a ^3}  \Delta_c  {\Omega_a ^2}  {\gamma_2 ^2}-64 {\Delta_a ^3}  \Delta_c  {\Omega_p ^4}+32 {\Delta_c ^2}  {\Delta_s ^2}  {\Omega_c ^2}  {\Omega_p ^2}\nonumber \\
	&-16 \Delta_a  \Delta_c  {\Omega_a ^2}  {\Omega_c ^2}  {\Omega_p ^2}+128 {\Delta_c }^6  {\Omega_p ^2}+48 {\Delta_c ^4}  {\Omega_p ^4}\nonumber\\
	&-32 {\Delta_c ^3}  \Delta_s  {\Omega_c ^4}-64 {\Delta_c ^2}  {\Delta_s ^2}  {\Omega_a ^2}  {\Omega_p ^2}-64 {\Delta_c ^2}  {\Delta_s ^2}  {\Omega_a ^2}  {\Omega_p ^2}\nonumber\\
	&-8 \Delta_c  \Delta_s  {\Omega_a ^4}  {\gamma_2 ^2}+8 {\Delta_s ^2}  {\Omega_a ^4}  {\Omega_p ^2}-768 {\Delta_a ^2}  {\Delta_c ^3}  \Delta_s  {\Omega_p ^2}.
\end{align}

\section*{Appendix C: proof of theorem \ref{opd}}
By performing the first-order Taypor approximation on (\ref{output_PD}) at $\Omega_{z,m}=\Omega_{y,m}$, we obtain the output of the PD as
\begin{equation}
	\setlength\abovedisplayskip{2.5pt}
	\setlength\belowdisplayskip{2.5pt}
	\label{C1}
\begin{aligned}
	{v_m} &= {v_m}({\Omega _{y,m}},t) + \frac{{\partial {v_m}({\Omega _{y,m}},t)}}{{\partial {\Omega_{z,m}}}}({\Omega _{z,m}} - {\Omega _{y,m}})\\
	&\mathop  \approx \limits^{(d)} 2\sqrt {G_{\text{LNA}}} \alpha \sqrt {{\mathcal{P}_{l,m}}{\mathcal{P}_m}({\Omega _{y,m}})} \cos ({\phi _{l,m}} - {\phi _{p,m}}({\Omega _{y,m}}))\\
	&- 2\frac{{\pi d}}{{{\lambda _p}}}\sqrt {G_{\text{LNA}}} \alpha \sqrt {{\mathcal{P}_{l,m}}{\mathcal{P}_m}({\Omega _{y,m}})} {{\tilde A}_m}\cos ({\phi _{l,m}} - {\phi _{p,m}}({\Omega _{y,m}})\\
	&+ {\psi _{p,m}}({\Omega _{y,m}}))({\Omega_{z,m}} - {\Omega _{y,m}})\\
	&= 2\sqrt {G_{\text{LNA}}} \alpha \sqrt {{\mathcal{P}_{l,m}}{\mathcal{P}_m}({\Omega _{y,m}})} [\cos ({\phi _{l,m}} - {\phi _{p,m}}({\Omega _{eff,m}}))\\
	&- \frac{{\pi d}}{{{\lambda _p}}}{{\tilde A}_m}\cos ({\phi _{l,m}} - {\phi _{p,m}}({\Omega _{y,m}}) + {\psi _{p,m}}({\Omega _{y,m}}))\\
	&\times \frac{1}{\sqrt{B}}\sum\limits_{i = 0}^{N-1} {{\Omega _{x,i,m}}} \cos (2\pi \Delta {f_{ni,m}}t + \Delta {\phi _{ni,m}})],
\end{aligned}
\end{equation}
where the derivation of $(d)$ can be found in the Appendix C of \cite{gongmodel}; ${{\tilde A}_m} = \sqrt {{\mathcal{I}^2}\{ {\chi _m}^\prime ({\Omega _{y,m}})\}  + {\mathcal{R}^2}\{ {\chi _m}^\prime ({\Omega _{y,m}})\} } $ and ${\psi _{p,m}} = \arccos \frac{{\mathcal{I}\{ {\chi _m}^\prime ({\Omega _{y,m}})\} }}{{{{\tilde A}_m}}}$. Based on the results in Appendix A, we have 
\begin{equation}
	\setlength\abovedisplayskip{2.7pt}
	\setlength\belowdisplayskip{2.7pt}
	\small
\begin{aligned}
\mathcal{R}\{ {\chi _m}^\prime \}  &= C\frac{{4{\overline{o}_1}{\Omega_{y,m}^3} + 2{\overline{o}_2}{\Omega_{y,m}}}}{{{\underline{o}_1}\Omega_{y,m}^4 + {\underline{o}_2}\Omega_{y,m}^2 + {\underline{o}_3}}}\\
&- C\frac{{({\overline{o}_1}{\Omega_{y,m}^4} + {\overline{o}_2}{\Omega_{y,m}^2} + {\overline{o}_3})(4{\underline{o}_1}{\Omega_{y,m}^3} + 2{\underline{o}_2}\Omega_{y,m})}}{{{{({\underline{o}_1}\Omega_{y,m}^4 + {\underline{o}_2}\Omega_{y,m}^2 + {\underline{o}_3})}^2}}},
\end{aligned}
\end{equation}
and
\begin{equation}
	\setlength\abovedisplayskip{2.7pt}
	\setlength\belowdisplayskip{2.7pt}
	\small
	\begin{aligned}
\mathcal{I}\{ {\chi _m}^\prime \}&= -C\frac{{4{\overline{o}_4}{\Omega_{y,m}^3} + 2{\overline{o}_5}{\Omega_{y,m}}}}{{{\underline{o}_1}\Omega_{y,m}^4 + {\underline{o}_2}\Omega_{y,m}^2 + {\underline{o}_3}}} \\
&+C\frac{{({\overline{o}_4}{\Omega_{y,m}^4} + {\overline{o}_5}{\Omega_{y,m}^2} + {\overline{o}_6})(4{\underline{o}_1}{\Omega_{y,m}^3} + 2{\underline{o}_2}\Omega_{y,m})}}{{{{({\underline{o}_1}\Omega_{y,m}^4 + {\underline{o}_2}\Omega_{y,m}^2 + {\underline{o}_3})}^2}}}.
\end{aligned}
\end{equation}
By further substituting ${\Omega _{x,i,m}} = \frac{{{\mu _{45}}}}{\hbar }\sqrt{\frac{2\mathcal{P}_{x,i,m}}{A_ec\epsilon_0}}$ into (\ref{C1}), we obtain Theorem \ref{opd}.

\section*{Appendix D: proof of theorem \ref{crb}}
Based on (\ref{fi}), the CRBs of AoA $\theta_k$ and distance $r_k$ can be expressed as shown at the top of this page. 
\begin{figure*}[ht] 
	\centering
	\begin{equation}
		\setlength\abovedisplayskip{2.7pt}
		\setlength\belowdisplayskip{2.7pt}
		\small
		\label{crb_general}
		\begin{aligned}
			\mathrm{CRB}({\theta _k})={[{{\bf{F}}^{ - 1}}]_{11}} &= \frac{{{\sigma ^2}}}{2}\frac{{\sum\limits_{i = 0}^{N-1} {\sum\limits_{n = 1}^J {{{\left\| {\frac{{\partial {{\bf{g}}_i}[n]}}{{\partial {r_k}}}} \right\|}^2}} } }}{{\sum\limits_{i = 0}^{N-1} {\sum\limits_{n = 1}^J {{{\left\| {\frac{{\partial {{\bf{g}}_i}[n]}}{{\partial {r_k}}}} \right\|}^2}} } \sum\limits_{i = 0}^{N-1} {\sum\limits_{n = 1}^J {{{\left\| {\frac{{\partial {{\bf{g}}_i}[n]}}{{\partial {\theta _k}}}} \right\|}^2}} }  - {{\left( {\sum\limits_{i = 0}^{N-1} {\sum\limits_{n = 1}^J {\frac{{\partial {{\bf{g}}_i}[n]}}{{\partial {\theta _k}}}{{\left( {\frac{{\partial {{\bf{g}}_i}[n]}}{{\partial {r_k}}}} \right)}^H}} } } \right)}^2}}},\\
			\mathrm{CRB}({r _k}) ={[{{\bf{F}}^{ - 1}}]_{22}}&= \frac{{{\sigma ^2}}}{2}\frac{{\sum\limits_{i = 0}^{N-1} {\sum\limits_{n = 1}^J {{{\left\| {\frac{{\partial {{\bf{g}}_i}[n]}}{{\partial {r_k}}}} \right\|}^2}} } }}{{\sum\limits_{i = 0}^{N-1} {\sum\limits_{n = 1}^J {{{\left\| {\frac{{\partial {{\bf{g}}_i}[n]}}{{\partial {r_k}}}} \right\|}^2}} } \sum\limits_{i = 0}^{N-1} {\sum\limits_{n = 1}^J {{{\left\| {\frac{{\partial {{\bf{g}}_i}[n]}}{{\partial {\theta _k}}}} \right\|}^2}} }  - {{\left( {\sum\limits_{i = 0}^{N-1} {\sum\limits_{n = 1}^J {\frac{{\partial {{\bf{g}}_i}[n]}}{{\partial {\theta _k}}}{{\left( {\frac{{\partial {{\bf{g}}_i}[n]}}{{\partial {r_k}}}} \right)}^H}} } } \right)}^2}}}.
		\end{aligned}
	\end{equation}
	\hrulefill
\end{figure*}
To further derive $\mathrm{CRB}({\theta _k})$ and $\mathrm{CRB}({r _k})$, we need to first calculate ${\frac{{\partial {{\bf{g}}_i}[n]}}{{\partial {r_k}}}}$\ and ${\frac{{\partial {{\bf{g}}_i}[n]}}{{\partial {\theta _k}}}}$. Recalling that ${[{{\bf{g}}_i}[n]]_m} = {\alpha _k}{\kappa _m}{e^{ - j\frac{4\pi }{c}i\Delta f r_k}} {e^{ - j \frac{2\pi}{\lambda_c} md\sin \theta_k }}s[n]$, we thus have
\begin{equation}
	\setlength\abovedisplayskip{2.7pt}
	\setlength\belowdisplayskip{2.7pt}
	\small
\begin{aligned}
	{\left[ {\frac{{\partial {{\bf{g}}_i}[n]}}{{\partial {\theta _k}}}} \right]_m} &= {\alpha _k}{\kappa _m}{r_{m,i}}\left( { - j\frac{{2\pi }}{\lambda_c}md\cos \theta_k } \right){a_{m,i}}s[n], \\
	{\left[ {\frac{{\partial {{\bf{g}}_i}[n]}}{{\partial {r_k}}}} \right]_m} &= {\alpha _k}{\kappa _m}\left( { - j\frac{{4\pi }}{c}i \Delta f} \right){r_{m,i}}{a_{m,i}}s[n].
\end{aligned}
\end{equation}
Therefore, we can compute $\sum_{n = 1}^J {{{\left\| {\frac{{\partial {{\bf{g}}_i}[n]}}{{\partial {\theta _k}}}} \right\|}^2}}$, $\sum_{n = 1}^J {{{\left\| {\frac{{\partial {{\bf{g}}_i}[n]}}{{\partial {r_k}}}} \right\|}^2}} $ and $\sum_{n = 1}^J {\frac{{\partial {{\bf{g}}_i}[n]}}{{\partial {\theta _k}}}{{\left( {\frac{{\partial {{\bf{g}}_i}[n]}}{{\partial {r_k}}}} \right)}^H}} $ in (\ref{crb_general}) as follows 
\begin{equation}
	\setlength\abovedisplayskip{2.7pt}
	\setlength\belowdisplayskip{2.7pt}
	\label{crb_t}
	\small
\begin{aligned}
	&\sum\limits_{n = 1}^J {{{\left\| {\frac{{\partial {{\bf{g}}_i}[n]}}{{\partial {\theta _k}}}} \right\|}^2}}  
	= \frac{{4{\pi ^2}}}{{{\lambda_c^2}}}J{{\cal P}_{s}}{\left| {{\alpha _k}} \right|^2}{d^2}{\cos ^2}{\theta _k}\sum\limits_{m = 1}^M {{{\left| {{\kappa _m}} \right|}^2}{m^2}},\\
	&\sum\limits_{n = 1}^J {{{\left\| {\frac{{\partial {{\bf{g}}_i}[n]}}{{\partial {r_k}}}} \right\|}^2}}=\frac{{16{\pi ^2}}}{{{c^2}}}J{{\cal P}_{s}}{\left| {{\alpha _k}} \right|^2}i^2 \Delta f^2\sum\limits_{m = 1}^M {{{\left| {{\kappa _m}} \right|}^2}},\\
	&\sum\limits_{n = 1}^J {\frac{{\partial {{\bf{g}}_i}[n]}}{{\partial {\theta _k}}}{{\left( {\frac{{\partial {{\bf{g}}_i}[n]}}{{\partial {r_k}}}} \right)}^H}}= \frac{{8{\pi ^2}}}{{{c\lambda_c}}}J{{\cal P}_{s}}{\left| {{\alpha _k}} \right|^2}d\cos {\theta _k} i \Delta f\sum\limits_{m = 1}^M {{{\left| {{\kappa _m}} \right|}^2}m}.
\end{aligned} 
\end{equation}
By substituting (\ref{crb_t}) into (\ref{crb_general}), we obtain the denominator of $\mathrm{CRB}(\theta_k)$ and $\mathrm{CRB}(r_k)$ as
\begin{align}
	\setlength\abovedisplayskip{2.7pt}
	\setlength\belowdisplayskip{2.7pt}
	\small
\label{denominator}
\begin{aligned}
	C_d
	&= \frac{{64{\pi ^4}}}{{{c^2\lambda_c^2}}}{J^2}{\cal P}_{s}^2{\left| {{\alpha _k}} \right|^4}{\Delta f^2d^2}{\cos ^2}{\theta _k}{(\sum\limits_{m = 1}^M {{{\left| {{\kappa _m}} \right|}^2}} )^2}\\
	&\times {N^2}(N + 1)\left( {\frac{{(2N + 1)}}{6}{\mu _2} - \frac{{N + 1}}{4}\mu _1^2} \right),
\end{aligned}
\end{align}
where ${\mu _1} \triangleq \frac{{\sum_{m = 1}^M {{{\left| {{\kappa _m}} \right|}^2}} m}}{{\sum_{m = 1}^M {{{\left| {{\kappa _m}} \right|}^2}} }}$ and ${\mu _2} \triangleq \frac{{\sum_{m = 1}^M {{{\left| {{\kappa _m}} \right|}^2}} {m^2}}}{{\sum_{m = 1}^M {{{\left| {{\kappa _m}} \right|}^2}} }}.$ By further substituting (\ref{denominator}) into (\ref{crb_general}), we arrive at the expression of CRBs of AoA and distance as 
\begin{align}
	\setlength\abovedisplayskip{2.7pt}
	\setlength\belowdisplayskip{2.7pt}
	\small
\begin{aligned}
	\mathrm{CRB}({\theta _k})
	&= \frac{{\lambda _c^2}}{{4{\pi ^2}J{{\overline {\mathcal{P}} }_{i,k}}{d^2}{{\cos }^2}{\theta _k}}}\\
	&\times \frac{{2N + 1}}{{N\sum\limits_{m = 1}^M {{{\left| {{\kappa _m}} \right|}^2}}((4N + 2){\mu _2} - 3(N + 1)\mu _1^2)}},
\end{aligned}
\end{align}
and
\begin{align}
	\setlength\abovedisplayskip{2.7pt}
	\setlength\belowdisplayskip{2.7pt}
	\small
\begin{aligned}
		\mathrm{CRB}({r_k}) 
		&= \frac{{{c^2}}}{{32{\pi ^2}J{{\overline {\mathcal{P}} }_{i,k}}\Delta {f^2}}}\\
		&\times \frac{{{\mu _2}}}{{{N}(N + 1)\sum\limits_{m = 1}^M {{{\left| {{\kappa _m}} \right|}^2}} \left( {\frac{{(2N + 1)}}{6}{\mu _2} - \frac{{N + 1}}{4}\mu _1^2} \right)}},
\end{aligned}
\end{align}
where ${\overline {\mathcal{P}} _{i,k}} \buildrel \Delta \over = \frac{{{{\left| {{\alpha _k}} \right|}^2}{{\mathcal{P}}_{s}}}}{{{\sigma _{w,i}^2}}}.$

\bibliographystyle{IEEEtran}
\bibliography{ref}

\end{document}